\documentclass[aps,pra,reprint,amsmath,amssymb,groupedaddress,showpacs]{revtex4-1}
\usepackage{float}
\usepackage{graphicx}%Include figure files
\usepackage{dcolumn} %Align table columns on decimal point
\usepackage{bm}      %bold math
\usepackage[bookmarks=false]{hyperref}%Hyperlinks
\usepackage[font={small}]{caption}    %FIXES FIGURE CAPTION PROBLEM
\hypersetup{
    unicode=false,          % non-Latin characters in Acrobat's bookmarks
    pdftoolbar=true,        % show Acrobat's toolbar?
    pdfmenubar=true,        % show Acrobat's menu?
    pdffitwindow=true,      % window fit to page when opened
    pdfstartview={},        % fits the width of the page to the window FitH
    pdftitle={Distinguishing Coherent States from Phase-Mixed Coherent States with Only a Variable Beam Splitter and Single-Photon Detector, In Principle},                 % title
    pdfauthor={Samuel R. Hedemann},% author
    pdfsubject={Quantum Optics},% subject of the document
    pdfcreator={},          % creator of the document
    pdfproducer={},         % producer of the document
    pdfkeywords={Coherent States,} {Coherent,} {Phase-Mixed,} {Phase-Mixed Coherent States,} {Variable Beam Splitter,} {Beam Splitter,} {Phase-Damping,} {Phase Diffusion,} {Laser,} {Single-Photon Detector,} {On-Off Detector,}  {Quantum Optics,} {Tomography}, % list of keywords
    pdfnewwindow=true,      % links in new window
    colorlinks=true,        % false: boxed links; true: colored links
    linkcolor=black,        % color of internal links (change box color with linkbordercolor)
    citecolor=black,        % color of links to bibliography
    filecolor=black,        % color of file links
    urlcolor=black          % color of external links
}%All links are clickable black text->least obtrusive, easy to find with mouse 
\newcommand{\Sec}[1]{\hyperref[sec:#1]{Sec.{\kern 2pt}\ref*{sec:#1}}}
\newcommand{\Section}[1]{\hyperref[sec:#1]{Section~\ref*{sec:#1}}}
\newcommand{\Fig}[2][]{\hyperref[fig:#2]{Fig.{\kern 2pt}\ref*{fig:#2}#1}}%#1=a,b,etc. is the optional argument, in [] at call. Ex: \Fig[c]{2} = Fig. 2c
\newcommand{\Figure}[2][]{\hyperref[fig:#2]{Figure~\ref*{fig:#2}#1}}%#1=a,b,etc. is the optional argument, in [] at call. Ex: \Fig[c]{2} = Fig. 2c
\newcommand{\App}[1]{\hyperref[sec:#1]{App.{\kern 2pt}\ref*{sec:#1}}}
\newcommand{\Appendix}[1]{\hyperref[sec:#1]{Appendix~\ref*{sec:#1}}}
\begin{document}
%*******************************************************************************
%                                   TITLE
\title{Distinguishing Coherent States from Phase-Mixed Coherent States with Only a Variable Beam Splitter and Single-Photon Detector, In Principle}
%*******************************************************************************
%*******************************************************************************
%                                  BYLINES
\author{Samuel R. Hedemann}
\affiliation{The Johns Hopkins University Applied Physics Laboratory, Laurel, MD 20723, USA}
\date{\today}
%*******************************************************************************
%*******************************************************************************
\begin{abstract}%                 0. ABSTRACT
It is generally assumed that on-off detectors with single-photon sensitivity cannot distinguish coherent states from phase-mixed coherent states without some form of quadrature-based tomography.  Here, we show that it is theoretically possible to distinguish these states without quadrature-based methods by comparing the vacuum probability of one output port of a variable beam splitter (VBS) that has two different phase-mixed coherent states as its inputs to the well-known case of dual coherent-state inputs.  As an application, a method is proposed to test whether a laser field is in a coherent state over a given time by simply measuring the VBS output populations with an on-off single-photon detector.  Unfortunately, timing limitations of present technology prohibit such simple tests, but the technique may become practical in the future.
\end{abstract}
%*******************************************************************************
%*******************************************************************************
%                             PACS & TITLE COMMAND
 %Interferometers
\maketitle
%*******************************************************************************
%*******************************************************************************
%                              I. INTRODUCTION
\section{\label{sec:I}Introduction}
Since optical systems offer a relatively affordable testbed for ideas in quantum mechanics, the action of a variable beam splitter (VBS) has been studied for a wide variety of input states of light.  In particular, papers like \cite[]{KSBK} have shown that the input state must be nonclassical for a VBS to produce entanglement in the output state, and have examined this proposition for a number of different inputs including Fock, coherent \cite[]{GLAU}, squeezed \cite[]{STOL,YUEN}, and thermal states. Others have investigated how to use beam splitters to filter out Fock states under certain special cases \cite[]{SRZ}. The VBS has even been investigated as a means to realize the displacement operator \cite[]{PARI}.

The motivation for studying the VBS is that it acts as a unitary operator that can be easily controlled, making it an attractive tool for realizations of quantum-optical computers, which is an active area of research, particularly in the subjects of linear quantum optical computing \cite[]{KMNR,PJF} and quantum-Zeno optical computing \cite[]{FJP}.

The motivations for developing a quantum computer \cite[]{FEYN} are well-documented, the chief reason being that quantum computers may be able to solve certain kinds of problems much more efficiently than classical computers.  However, the production and control of pure quantum states is very challenging, and it is often difficult to verify quantum features such as superposition that are essential to quantum algorithms.

Here, we seek a method of verifying the purity of a coherent state subject only to phase-damping noise, without using full tomography.  Specifically, since a laser field is in a pure coherent state over small enough times \cite[]{ZPKB}, we ask: \textit{can we verify that a laser field is still in a pure coherent state over a certain time using only linear optics and an on-off detector?}  This is especially relevant since lasers are vital for control in many implementations of quantum computation.  This question was in fact addressed by \cite[]{MDB} in 1999, however their method relies on quadrature measurements, which we seek to avoid here.

Over time, a laser field's phase does a random-walk, yielding a phase-damping effect which reduces all of the off-diagonal elements of the density matrix to zero in the Fock photon-number basis, leaving the probabilities unchanged, the result of which is called a \textit{phase-mixed coherent state}.  Since these probabilities are the same as those of the original pure coherent state, we cannot test the purity of the laser field by checking its probabilities.  Thus, in order to develop a simple method for distinguishing these two states, we now review their mathematical descriptions as well as those of the VBS.

Recall that Glauber's coherent states \cite[]{GLAU} are
%===============================================================================
\begin{equation}%                  Equation 1
|\alpha \rangle  = \sqrt {e^{ - |\alpha |^2 } } \sum\limits_{n = 0}^\infty  {{\textstyle{{\alpha ^n } \over {\sqrt {n!} }}}|n\rangle },
\label{eq:1}
\end{equation}
%===============================================================================
where $\{|n\rangle\}$ is the Fock basis, and $\alpha  \equiv |\alpha |e^{i\varphi }$ is complex, with real phase angle $\varphi$.  The density operator of (\ref{eq:1}) is
%===============================================================================
\begin{equation}%                  Equation 2
\rho _\alpha   \equiv |\alpha \rangle \langle \alpha | = e^{ - |\alpha |^2 } \sum\limits_{n = 0}^\infty  {\sum\limits_{m = 0}^\infty  {{\textstyle{{\alpha ^n } \over {\sqrt {n!} }}}{\textstyle{{{\alpha ^*} ^m } \over {\sqrt {m!} }}}|n\rangle \langle m|} }.
\label{eq:2}
\end{equation}
%===============================================================================
Although coherent states exhibit many classical features, they are still quantum in the sense of being pure states possessing superposition.  The truly classical counterparts to these states are the phase-mixed coherent states,
%===============================================================================
\begin{equation}%                  Equation 3
\rho _{\alpha _{\overline \varphi } }  \equiv {\textstyle{1 \over {2\pi }}}\int_0^{2\pi } \!\!\!\!\!{\rho_{\alpha}d\varphi }  = e^{ - |\alpha |^2 } \sum\limits_{n = 0}^\infty  {{\textstyle{{|\alpha |^{2n} } \over {n!}}}|n\rangle \langle n|},
\label{eq:3}
\end{equation}
%===============================================================================
which we shall call \textit{Poisson states}, where their classicality comes from being diagonal in the Fock basis, meaning that these mixed states are the results of coherent states that have been fully phase-damped.

\begin{figure}[H]%                   FIGURE 1
\centering
\includegraphics[width=0.99\linewidth]{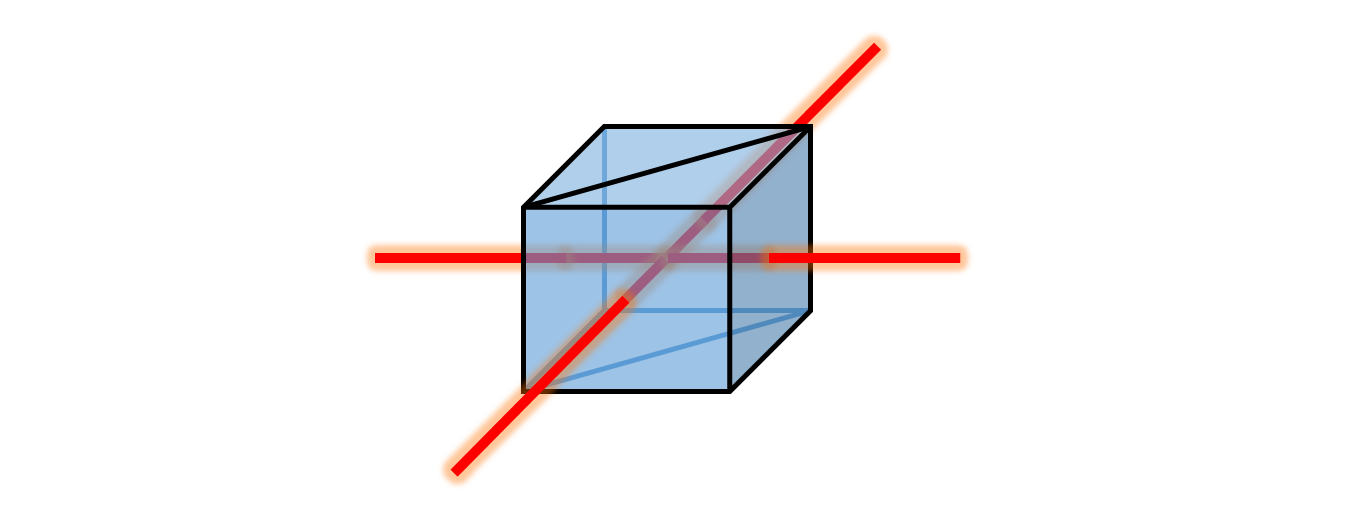}%Base name of figure file here.
\vspace{-12pt}
\setlength{\unitlength}{0.01\linewidth}
\begin{picture}(100,0)
\put(40.3,30.3){\normalsize VBS}
\put(58.8,8.8){\normalsize $B_{\theta}$}
\put(20,18){\normalsize $|\alpha\rangle$}
\put(26,1.8){\normalsize $|\beta\rangle$}
\put(68,33.5){\normalsize $| - \alpha s_\theta + \beta c_\theta\rangle$}
\put(73.8,17.8){\normalsize $|\alpha c_\theta + \beta s_\theta\rangle$}
\end{picture}
\caption[]{(color online) Simplified depiction of a variable beam splitter (VBS) with two different pure coherent-state inputs $|\alpha\rangle$ and $|\beta\rangle$.  The unitary operator of the VBS is $B_{\theta}$, where the transmittivity $T$ is related to parameter $\theta$ by \smash{$T\!=\!\text{cos}^2(\theta)\equiv c_{\theta}^{2}$}.}
\label{fig:1}
\end{figure}
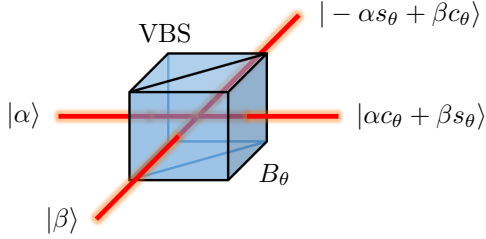

Recall also that \smash{$B_\theta  \equiv e^{\theta (a^{\dag}  b - ab^{\dag}  )}$} is the unitary operator of a VBS (without a second phase shifter), where annihilation operators $a$ and $b$ obey $[a,a^{\dag}  ] = 1$ and $[b,b^{\dag}  ] = 1$ where $[A,B]\equiv AB-BA$, and correspond to different input ports of the VBS \cite[]{NiCh}. The action of $B_\theta$ on $a$ and $b$ then gives the output operators as
%===============================================================================
\begin{equation}%                  Equation 4
\begin{array}{*{20}l}
   {a'} &\!\! {\equiv B_\theta  a B_\theta  ^{\dag}  } &\!\! { = a c_\theta   - b s_\theta  }  \\
   {b'} &\!\! {\equiv B_\theta  b B_\theta  ^{\dag}  } &\!\! { = a s_\theta   + b c_\theta  },  \\
\end{array}
\label{eq:4}
\end{equation}
%===============================================================================
where $c_\theta   \equiv \cos (\theta )$ and $s_\theta   \equiv \sin (\theta )$.  Then, using the fact that \smash{$|n\rangle  = {\textstyle{{(a^{\dag}  )^n } \over {\sqrt {n!} }}}|0\rangle$}, and letting $|A\rangle |B\rangle  \equiv |A\rangle  \otimes |B\rangle$, where $a$ operates on the left ket and $b$ operates on the right ket, and recalling the unitary displacement operator \smash{$D(\alpha ) \equiv e^{\alpha a^ {\dag}   - \alpha ^* a}$} such that $|\alpha \rangle  = D(\alpha )|0\rangle$, and using the facts that \smash{$|0\rangle  = e^{ - \alpha ^* a} |0\rangle$} and \smash{$e^{\alpha a^{\dag}  } e^{ - \alpha ^* a}  = \sqrt {e^{|\alpha |^2 } } e^{\alpha a^{\dag}   - \alpha ^* a}$}, we get the well-known result of applying a VBS to dual pure-coherent inputs as
%===============================================================================
\begin{equation}%                  Equation 5
B_\theta  |\alpha \rangle |\beta \rangle  = |\alpha c_\theta   + \beta s_\theta  \rangle | - \alpha s_\theta   + \beta c_\theta  \rangle,
\label{eq:5}
\end{equation}
%===============================================================================
where $\beta$ is also any complex number.  Thus, a VBS transforms the product of pure coherent states to a new product of pure coherent states, whose complex parameters are a unitary transformation of the complex input parameters, as depicted in \Fig{1}.
%_______________________________________________________________________________
%_______________________________________________________________________________

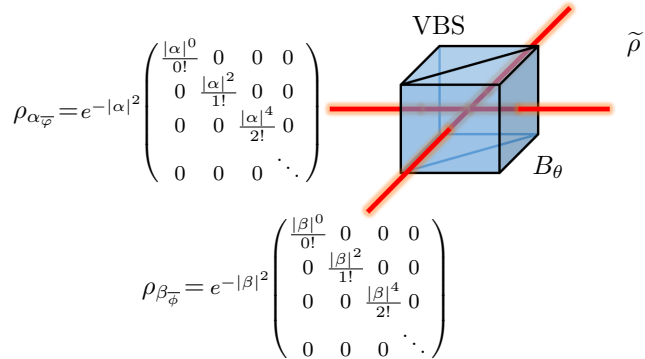
\begin{figure}[H]%                   FIGURE 2
\centering
\includegraphics[width=0.99\linewidth]{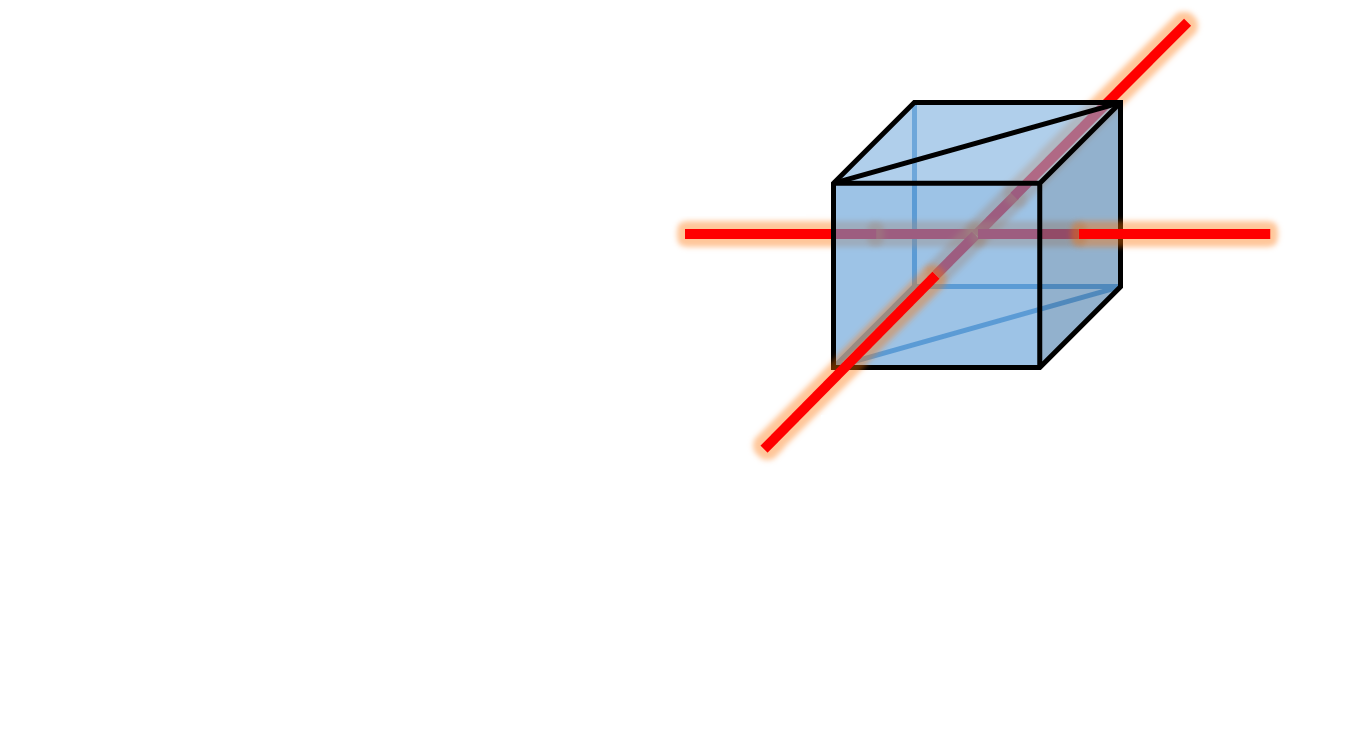}%Base name of figure file here.
\vspace{-12pt}%fixes gap caused by picture zero-height picture environment below
\setlength{\unitlength}{0.01\linewidth}
\begin{picture}(100,0)%0 height keeps from interfering with pic height, but gap
\put(2.1,36.1){\normalsize $\rho _{\alpha _{\overline \varphi } }\!=$}
\put(13.1,36.1){\footnotesize $\!e^{-|\alpha|^{2}}\!\!\left(\! {\!\begin{array}{*{20}c}
   {{\textstyle{{|\alpha |^0 } \over {0!}}}} &\!\!\!\!\! 0 &\!\!\!\!\! 0 &\!\!\!\!\!  0   \\
   0 &\!\!\!\!\! {{\textstyle{{|\alpha |^2 } \over {1!}}}} &\!\!\!\!\! 0 &\!\!\!\!\!  0   \\
   0 &\!\!\!\!\! 0 &\!\!\!\!\! {{\textstyle{{|\alpha |^4 } \over {2!}}}} &\!\!\!\!\!  0   \\
    0  &\!\!\!\!\!  0  &\!\!\!\!\!  0  &\!\!\!\!\!  \ddots   \\
\end{array}}\! \right)$}
\put(21.9,9){\normalsize $\rho _{\beta _{\overline \phi } }\!=$}
\put(33,9){\footnotesize $\!e^{-|\beta|^{2}}\!\!\left(\! {\!\begin{array}{*{20}c}
   {{\textstyle{{|\beta |^0 } \over {0!}}}} &\!\!\!\!\! 0 &\!\!\!\!\! 0 &\!\!\!\!\!  0   \\
   0 &\!\!\!\!\! {{\textstyle{{|\beta |^2 } \over {1!}}}} &\!\!\!\!\! 0 &\!\!\!\!\!  0   \\
   0 &\!\!\!\!\! 0 &\!\!\!\!\! {{\textstyle{{|\beta |^4 } \over {2!}}}} &\!\!\!\!\!  0   \\
    0  &\!\!\!\!\!  0  &\!\!\!\!\!  0  &\!\!\!\!\!  \ddots   \\
\end{array}}\! \right)$}
\put(96.5,46.5){\normalsize $\widetilde{\rho}$}
\put(63.5,49){\normalsize VBS}
\put(82,27.5){\normalsize $B_{\theta}$}
\end{picture}%see \vector and \framebox
\caption[]{(color online) Simplified depiction of a VBS with two different phase-mixed coherent states \smash{$\rho _{\alpha _{\overline \varphi } }$} and \smash{$\rho _{\beta _{\overline \phi } }$} (Poisson states) as defined in (\ref{eq:3}).  The output beams comprise a single system in state \smash{${\widetilde{\rho}}\equiv B_\theta (\rho _{\alpha _{\overline \varphi } }  \otimes \rho _{\beta _{\overline \phi } } )B_\theta ^{\dag}$}.}
\label{fig:2}
\end{figure}

The first question we wish to answer here is, \textit{what is the output of a VBS acting on two different Poisson states, and can we do anything useful with it?}.  That is, given these two very classical input states, can we distinguish them from the coherent-state case to verify nonclassical properties such as superposition using only classical measurements of probability with an on-off detector?  The situation of interest is shown in \Fig{2}.
%_______________________________________________________________________________
%_______________________________________________________________________________

In particular, since single-photon detectors are the most widely available detection devices, we will focus on the vacuum probability, the difference of that from 1 being the nonvacuum probability, meaning the probability of getting any pulse at all from an ideal on-off single-photon detector.  For the case of pure coherent inputs from (\ref{eq:5}), tracing over subsystem $2$, the vacuum probability is \smash{$p_0^{(1)}  \equiv \langle 0|\text{tr}_2 (B_{\theta}(\rho _{\alpha}^{(1)}  \otimes \rho _{\beta}^{(2)}) B_{\theta}^{\dag})|0\rangle$}, yielding
%===============================================================================
\begin{equation}%                  Equation 6
p_0^{(1)}  = e^{ - |\alpha c_\theta   + \beta s_\theta  |^2 }.
\label{eq:6}
\end{equation}
%===============================================================================

Now we will discuss the analogs of (\ref{eq:5}) and (\ref{eq:6}) for the case of the diagonal input of Poisson states.
%                                  END of I
%*******************************************************************************
%*******************************************************************************
%         II. OUTPUT OF VARIABLE BEAM SPLITTER ON TWO POISSON STATES
\section{\label{sec:II}Output of Variable Beam Splitter on Two Poisson States}
Using the Poisson states of (\ref{eq:3}), we use \smash{$|n\rangle  = {\textstyle{{(a^{\dag}  )^n } \over {\sqrt {n!} }}}|0\rangle$} with (\ref{eq:4}) and its Hermitian conjugates, along with the binomial expansion \smash{$(a + b)^x  = \sum\nolimits_{y = 0}^x {{\textstyle{{x!} \over {\left( {x - y} \right)!y!}}}a^{x - y} b^y }$} to obtain the general output state \smash{$\widetilde{\rho}\equiv B_\theta  (\rho _{\alpha _{\overline \varphi } }  \otimes \rho _{\beta _{\overline \phi } } )B_\theta  ^ {\dag}$} of a VBS given two different Poisson input states as
%===============================================================================
\begin{equation}%                  Equation 7
\begin{array}{*{20}l}
   {\widetilde{\rho} =} &\!\! {e^{ - |\alpha |^2  - |\beta |^2 } \mathop \sum \limits_{n = 0}^\infty  \mathop \sum \limits_{m = 0}^\infty  \left( {{\textstyle{{(|\alpha |c_\theta )^{n}  } \over {n!}}}} \right)^{2}\! \left( {{\textstyle{{(|\beta |s_\theta )^{m} } \over {m!}}}} \right)^{2} }  \\
   {} &\!\! {\times \sum\limits_{q = 0}^n {\sum\limits_{r = 0}^m {{\textstyle{{n!\sqrt {\left( {n + m - q - r} \right)!} } \over {\left( {n - q} \right)!q!}}}{\textstyle{{m!\sqrt {(q + r)!} } \over {\left( {m - r} \right)!r!}}}\left( {{\textstyle{{ - s_\theta  } \over {c_\theta  }}}} \right)^q \left( {{\textstyle{{c_\theta  } \over {s_\theta  }}}} \right)^r } }}  \\
   {} &\!\! {\times \sum\limits_{s = 0}^n {\sum\limits_{t = 0}^m {{\textstyle{{n!\sqrt {\left( {n + m - s - t} \right)!} } \over {\left( {n - s} \right)!s!}}}{\textstyle{{m!\sqrt {\left( {s + t} \right)!} } \over {\left( {m - t} \right)!t!}}}\left( {{\textstyle{{ - s_\theta  } \over {c_\theta  }}}} \right)^s \left( {{\textstyle{{c_\theta  } \over {s_\theta  }}}} \right)^t } }}  \\
   {} &\!\! {\times |n + m - q - r\rangle \langle n + m - s - t| \otimes |q + r\rangle \langle s + t|.}  \\
\end{array}
\label{eq:7}
\end{equation}
%===============================================================================
This can be viewed as a mixture of many pure states of distinct total photon number $N\equiv n+m$. In each total-photon subspace of the final state, there is superposition, which was created by the VBS acting on the Fock states of the input state.

However, the main usefulness of (\ref{eq:7}) is as a starting point for finding other quantities of interest.  In particular, we will now show that the vacuum probability of the first output port takes on a simple form.  Defining \smash{$\widetilde{p}_0^{(1)}  \equiv \langle 0|\text{tr}_2 (B_\theta  (\rho _{\alpha _{\overline \varphi } }  \otimes \rho _{\beta _{\overline \phi } } )B_\theta  ^{\dag}  )|0\rangle$}, we obtain
%===============================================================================
\begin{equation}%                  Equation 8
\widetilde{p}_0^{(1)}  = e^{ - |\alpha c_\theta |^2  - |\beta s_\theta |^2 } I_0 (|\alpha ||\beta |s_{2\theta } ),
\label{eq:8}
\end{equation}
%===============================================================================
as shown in \App{App.A}, where \smash{$I_0 (x) \equiv \sum\nolimits_{n = 0}^\infty  {{\textstyle{{(x^2 /4)^n } \over {(n!)^2 }}}} $} is the modified Bessel function of the first kind, order zero, and again, \smash{$c_\theta   \equiv \cos (\theta )$} and \smash{$s_\theta   \equiv \sin (\theta )$}.

Immediately, we see that (\ref{eq:8}) is different from (\ref{eq:6}), revealing that there \textit{is} a difference in the VBS output probabilities accessible to an on-off detector.  Therefore we now investigate these results visually to see if they are different enough to be useful in practice.
%-------------------------------------------------------------------------------
%  II.A. Comparison of Coherent Inputs and Poisson Inputs to a Variable Beam Splitter
\subsection{\label{sec:II.A}Comparison of Coherent Inputs and Poisson Inputs to a Variable Beam Splitter}
Since the output pulses of an ideal typical single-photon detector describe \textit{nonvacuum} events, meaning collapses of the field state onto outcomes other than the vacuum state, then here we will study the nonvacuum probabilities \smash{$p\equiv 1-p_{0}^{(1)}$} and \smash{$\widetilde{p}\equiv 1-\widetilde{p}_{0}^{(1)}$}.  Our goal here is to visually check whether there is any noticeable and useful difference between $p$ and \smash{$\widetilde{p}$}.

\begin{figure}[H]%                   FIGURE 3
\centering
\includegraphics[width=0.99\linewidth]{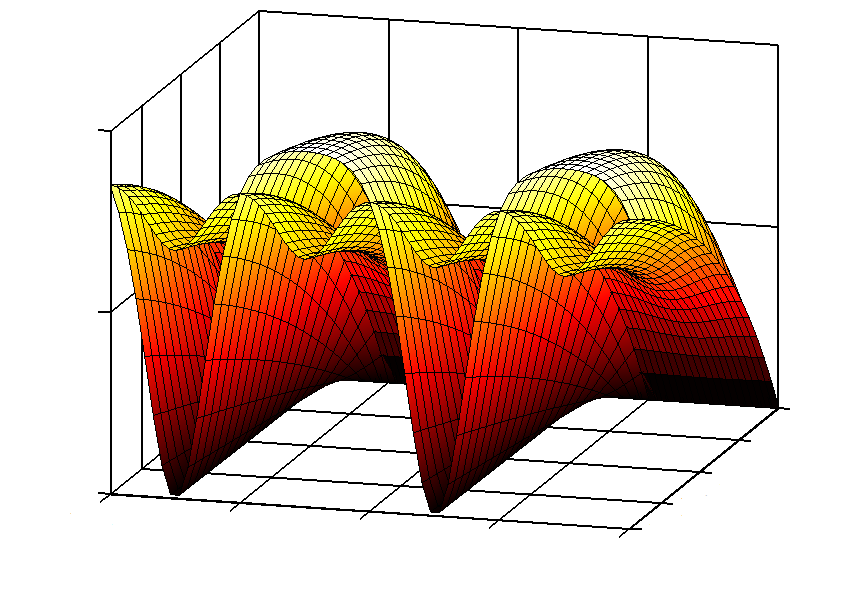}%Base name of figure file here.
\vspace{-12pt}
\setlength{\unitlength}{0.01\linewidth}
\begin{picture}(100,0)
\put(1.7,45){\normalsize $p,\widetilde{p}$}
\put(36,1.5){\normalsize $\theta\;[\text{deg}]$}
\put(90,11){\normalsize $|\alpha|^2$}
\put(36.5,61){\normalsize $p$}
\put(38,60.3){\vector(1,-4){1.1}}
\put(52.7,52.8){\normalsize $\widetilde{p}$}
\put(54.3,52.1){\vector(1,-4){1.6}}
%Zaxis: p (probability)
\put(6.1,55.5){$1.0$}
\put(6,33.8){$0.5$}
\put(6,12.8){$0.0$}
%Yaxis: \theta (VBS angle)
\put(8,8.9){$360$}
\put(23,7.8){$270$}
\put(38.2,6.8){$180$}
\put(53.9,5.6){$90$}
\put(70.3,4.8){$0$}
%Xaxis: |\alpha|^2 (mean photon number of pure coherent state)
\put(93,21.9){$0.0$}
\put(88.6,18){$0.5$}
\put(83.8,14.3){$1.0$}
\put(79.5,10.8){$1.5$}
\put(75.8,7.7){$1.9$}
\end{picture}
\caption[]{(color online) Nonvacuum probabilities \smash{$p\equiv 1-p_{0}^{(1)}=$} \smash{$1-e^{-|\alpha|^{2}(1+s_{2\theta})}$} and \smash{$\widetilde{p}\equiv 1-\widetilde{p}_{0}^{(1)}=1-e^{ - |\alpha |^2} I_0 (|\alpha |^{2}s_{2\theta } )$} of the state at VBS output-port $1$ vs.~VBS angle $\theta$ and the square magnitude of coherent state parameter \smash{$\alpha\equiv|\alpha|e^{i\varphi}$} where $\varphi=0$ and $\beta=\alpha$, for the cases of both VBS inputs being pure coherent states and both VBS inputs being diagonal phase-mixed coherent states (Poisson states).  Although $|\alpha|^{2},|\beta|^{2}\in[0,\infty)$, this plot only shows the case where $|\alpha|^{2}=|\beta|^{2}\in[0,-\text{ln}(1-p_{*})]$ where $p_{*}=0.85$, chosen for the view it offers. The coherent-generated surface $p$ is the one with only two periods of repetition with $\theta$, while the Poisson-generated surface \smash{$\widetilde{p}$} has four periods with $\theta$.}
\label{fig:3}
\end{figure}
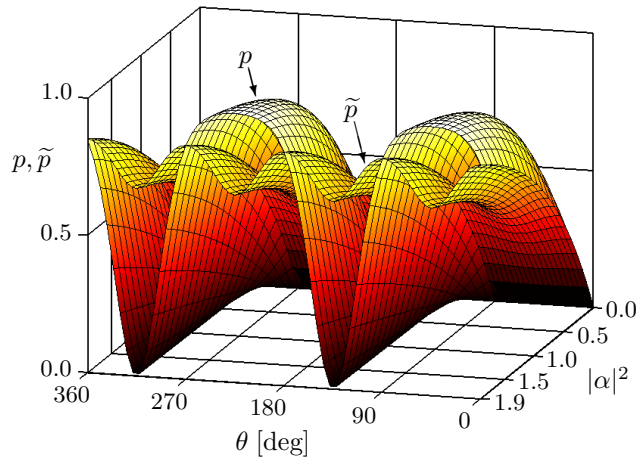

In fact, there \textit{is} a noticeable difference, as seen in \Fig{3}, which uses the simplifying condition that $|\alpha|^{2}=|\beta|^{2}$, which are the mean photon numbers of each input beam for both the coherent case and the Poisson case.
%_______________________________________________________________________________
%_______________________________________________________________________________

As \Fig{3} shows, the periodicity of the Poisson-generated surface \smash{$\widetilde{p}$} is twice that of the coherent-generated surface $p$ as $\theta$ is varied over a full period.  The extreme cases when $|\alpha|^{2} =0$ and  $|\alpha|^{2} =\infty$ cause \smash{$p=\widetilde{p}=0$} and \smash{$p=\widetilde{p}=1$} respectively, and therefore are not as useful as the region where $|\alpha|^{2}$ has low nonzero values.  In particular, the case corresponding to $p_{*}=0.85$ gives $|\alpha|^{2} =-\text{ln}(1-p_{*})\approx 1.9$ which is a mean photon number that approximately yields the most pronounced differences between the behaviors of $p$ and \smash{$\widetilde{p}$}.

Thus, while there are some points of intersection, as seen in \Fig{3}, it is permissible to suppose that, adjusting to get $|\alpha|^{2} =|\beta|^{2}\approx 1.9$, one could then vary the VBS angle $\theta$ and collect data to see which curve is obtained, with the time-interval of interest being defined by the time-window over which the detector is allowed to operate during each Bernoulli trial of a binomial experiment.  We will discuss this in more detail later. 

In practice, one can let $|\alpha|^{2}$ and $|\beta|^{2}$ be different and use them as fitting parameters for the data. However, one important limiting case should be avoided; one must apply nonvacuum states to \textit{both} inputs of the VBS in order to obtain different detection probabilities in one of the output ports.  To see why, putting $\beta=0$ into (\ref{eq:6}) and (\ref{eq:8}) produces \smash{$p_0^{(1)}  = e^{ - |\alpha c_\theta |^2 }$} and \smash{$\widetilde{p}_0^{(1)}  = e^{ - |\alpha c_\theta |^2 } I_0 (0)=e^{ - |\alpha c_\theta |^2 }$}, making the on-off output statistics identical, and therefore useless to compare.

Note that nonvacuum probability $p$ only relates to $|\alpha|^{2}$ as $|\alpha|^{2} =-\text{ln}(1-p)$ for \textit{both} coherent and Poisson states if $p$ is measured for each single beam \textit{prior} to the VBS.  Therefore, $p_{*}$ is a pre-VBS quantity.

Before using these results to develop a test, we now take a closer look to make sure that no problems arise due to the phase shifts in the coherent-input case.
%                                 End of II.A
%-------------------------------------------------------------------------------
%-------------------------------------------------------------------------------
%    II.B. Verification that Phase Shifts Do Not Cause the Coherent Case to Look Like the Poisson Case
\subsection{\label{sec:II.B}Verification that Phase Shifts Do Not Cause the Coherent Case to Look Like the Poisson Case}
Here, we wish to look at the cases where the single-beam nonvacuum output probabilities $p$ and $\widetilde{p}$ are the most different, and check that they are reliably distinguishable, which would then license the creation of a simple test to determine whether the input states were coherent states or Poisson states.

In this section, we limit ourselves to the case where $|\alpha|^{2} =|\beta|^{2}\approx 1.9$, and we assume that the input states have been created from identical monochromatic laser beams (the preparation of which we will discuss in more detail soon), with a stable phase relationship.

Then, since small and unavoidable changes in path length between the two beams will occur within the VBS, if we use one beam as a reference, the other will have some unknown classical phase shift with respect to that beam.  Thus, here we seek to determine whether there are any dangerous classical phase shifts that would make the nonvacuum probabilities of the two cases identical.  If all classical phase shifts produce curves distinguishable from the Poisson case, then we have a robust simple test for verifying the purity of the initial laser field state.

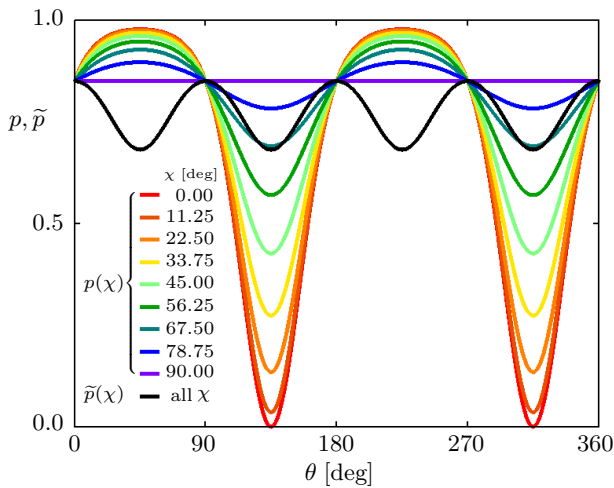
\begin{figure}[H]%                   FIGURE 4
\centering
\includegraphics[width=0.99\linewidth]{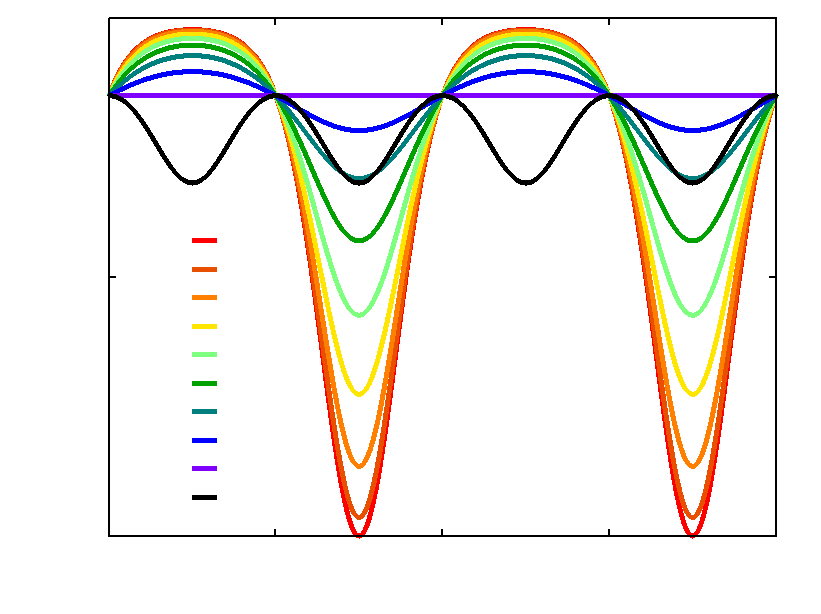}%Base name of figure file here.
\vspace{-12pt}
\setlength{\unitlength}{0.01\linewidth}
\begin{picture}(100,0)
\put(3.5,55){$p,\widetilde{p}$}
\put(49,0.5){$\theta\;[\text{deg}]$}
%Yaxis:
\put(7.1,70.5){$1.0$}
\put(7.1,39){$0.5$}
\put(7.1,8.2){$0.0$}
%Xaxis:
\put(12.6,5){$0$}
\put(31.5,5){$90$}
\put(51.0,5){$180$}
\put(71.0,5){$270$}
\put(90.9,5){$360$}
%Legend:
\put(15,30.0){\scriptsize $p(\chi)$}
\put(21.2,30.3){\tiny $\mathbf{\left\{ {\begin{array}{*{20}c}
   {}  \\
   {}  \\
   {}  \\
   {}  \\
   {}  \\
   {}  \\
   {}  \\
   {}  \\
   {}  \\
   {}  \\
   {}  \\
   {}  \\
\end{array}} \right.}$}
\put(27.3,47){\tiny $\chi\;[\text{deg}]$}
\put(29.1,43.5){\scriptsize  $0.00$}
\put(27.6,40.2){\scriptsize $11.25$}
\put(27.6,36.8){\scriptsize $22.50$}
\put(27.6,33.4){\scriptsize $33.75$}
\put(27.6,30.0){\scriptsize $45.00$}
\put(27.6,26.5){\scriptsize $56.25$}
\put(27.6,23.0){\scriptsize $67.50$}
\put(27.6,19.5){\scriptsize $78.75$}
\put(27.6,16.3){\scriptsize $90.00$}
\put(15,13.0){\scriptsize $\widetilde{p}(\chi)$}
\put(28.3,12.7){\scriptsize $\text{all}$}
\put(32.4,13.2){\scriptsize $\chi$}
\end{picture}
\caption[]{(color online) Nonvacuum probabilities of one output beam of a VBS where \smash{$p\equiv 1-p_{0}^{(1)}=1-e^{-|\alpha|^{2}(1+s_{2\theta}c_{\chi})}$} is for the dual-coherent-input case, and \smash{$\widetilde{p}\equiv 1-\widetilde{p}_{0}^{(1)}=$} \smash{$1-e^{ - |\alpha |^2} I_0 (|\alpha |^{2}s_{2\theta } )$} is for the dual-Poisson-input case, and $\theta$ is the VBS angle.  This plot treats the case where $|\alpha|^2 =|\beta|^2 =-\text{ln}(1-p_{*})$, where $p_{*}=0.85$ causes the most visibly different behaviors between $p$ and \smash{$\widetilde{p}$}.  This figure examines the effect of changing the relative phase angle $\chi\equiv \phi-\varphi$ between $\alpha$ and $\beta$, taking $\text{arg}(\alpha)=\varphi\equiv 0$ as the reference so $\chi=\phi=\text{arg}(\beta)$. Note that \smash{$\widetilde{p}$} is unaffected by $\chi$, and has twice as many periods with $\theta$ than $p$.  Most importantly, this plot shows that, under these conditions and if measurements are separated \textit{at most} by $45^{\circ}$, we can reliably distinguish $p$ from \smash{$\widetilde{p}$}, as long as $\theta$ is varied by at least $180^{\circ}$, so that we see enough of the curves to avoid the similarity of \smash{$p(67.5^{\circ})$} to \smash{$\widetilde{p}$} when \smash{$\theta\in[90^{\circ},180^{\circ}]\;\text{and}\;[270^{\circ},360^{\circ}]$}.}
\label{fig:4}
\end{figure}

To examine the effects of classical phase shift, \Fig{4} examines the case where $|\alpha|^{2} =|\beta|^{2}\approx 1.9$ for various differences between coherent-state phase angles $\chi\equiv\text{arg}(\beta)-\text{arg}(\alpha)=\phi-\varphi$, where we arbitrarily assign $\varphi=0$ as a reference.  Although these are phase angles of coherent states which are quantum states, it is well-known that the expectation value of the electric field operator for a field in a single-mode coherent state produces a classical electric plane-wave field with classical phase shift equal to the phase angle of the coherent state, which is why we refer to $\chi$ as a classical phase shift.

As \Fig{4} shows, under the specified conditions, the coherent-generated curves $p$ are reliably distinguishable from the Poisson-generated curve $\widetilde{p}$ for all values of $\chi$ shown, where we omit other values because $\chi\in[90^{\circ},180^{\circ}]$ simply produces a horizontal translation of the curves shown resulting in a qualitatively identical situation, and then $\chi\in[180^{\circ},360^{\circ}]$ simply causes the reverse sequence of curves.
%_______________________________________________________________________________
%_______________________________________________________________________________

The only case in \Fig{4} where the curves are even remotely similar is when $\chi=67.5^{\circ}$ since $p$ and \smash{$\widetilde{p}$} have similar shapes for two quadrants of $\theta$, but then, for the other two quadrants, the curves conveniently spread away with different curvatures and are easily distinguished.

Therefore, it does not matter what the classical phase shift $\chi$ is; as long as one varies the VBS angle $\theta$ over at least a half of its full period, then it should be possible to get enough data to say which curve is being observed.

Now that we have verified that the two different input state-pairs can yield distinguishable results from a \textit{theoretical} point of view, we need to briefly consider some practical concerns for implementing this in a lab, in particular, \textit{how do we prepare these input states?}
%                                End of II.B
%-------------------------------------------------------------------------------
%-------------------------------------------------------------------------------
%          II.C. Appraisal of Various Input-State Preparations
\subsection{\label{sec:II.C}Appraisal of Various Input-State Preparations}
Here we look at two simple ways of preparing input states to the VBS and discuss their merits and faults relative to the theoretical results we have just developed.  The first seems the safest, but is found to fail completely, while the second \textit{could} yield a robust test of laser field purity if practical obstacles can be overcome.
%...............................................................................
%                II.C.1 Single Laser Beam Split into Two Beams (Fails)
\subsubsection{\label{sec:II.C.1}Single Laser Beam Split into Two Beams (Fails)}
\begin{figure}[H]%                   FIGURE 5
\centering
\includegraphics[width=0.99\linewidth]{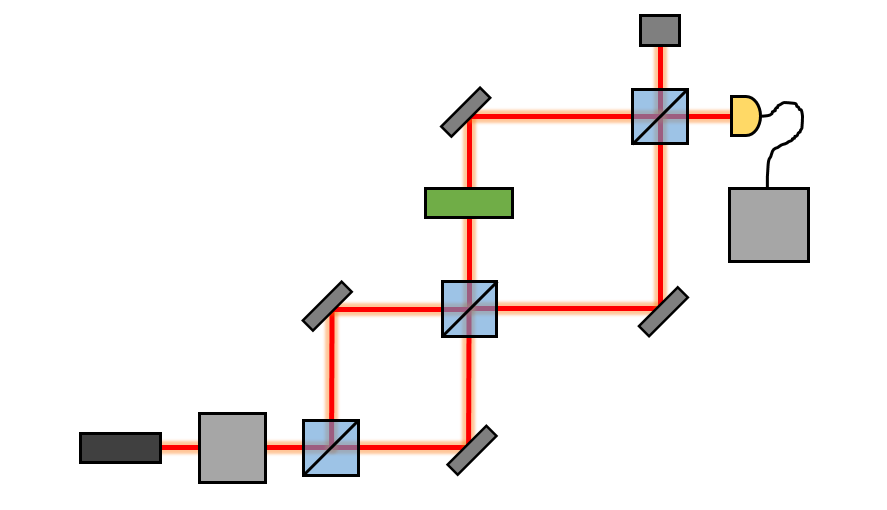}%Base name of figure file here.
\vspace{-12pt}
\setlength{\unitlength}{0.01\linewidth}
\begin{picture}(100,0)
\put(9.8,9.3){\normalsize $\text{laser}$}
\put(24.8,5){\normalsize $\text{A}$}
\put(41,9.3){\normalsize $\text{B}_1$}
\put(54,1.5){\normalsize $\text{M}$}
\put(32.1,22.9){\normalsize $\text{M}$}
\put(47,16){\normalsize $\rho$}
\put(56.6,24.8){\normalsize $\text{B}_2$}
\put(47.5,45){\normalsize $\text{M}$}
\put(75.5,16.9){\normalsize $\text{M}$}
\put(40.9,32.3){\normalsize $\text{PS}$}
\put(59.2,32.3){\normalsize $\theta$}
\put(66.3,37.4){\normalsize $\text{B}_3$}
\put(64.3,51.5){\normalsize $\text{BD}$}
\put(84.7,30){\normalsize $\text{E}$}
\put(81.6,46.7){\normalsize $\text{D}$}
\end{picture}
\caption[]{(color online) Schematic of a na{\"i}ve realization to test for laser field state purity. The laser beam under test enters attenuator A, resulting in mean photon number we can define as \smash{$|\sqrt{2}\alpha|^2$}, and is split by $50\!:\!50$ beam splitter $\text{B}_1$ creating joint-beam state $\rho$, which is sent by mirrors M to the variable beam splitter (VBS) which is a Mach-Zehnder interferometer starting at $\text{B}_2$.  The VBS angle $\theta$ is realized by a phase-shifter PS, and the VBS ends with $\text{B}_3$. Finally, one output beam goes to beam-dump BD, while the other output beam goes to detector D, the output of which is processed by electronics E.  Although this method ensures stable relative phase for the coherent-input case, it fails to produce a product of Poisson inputs at $\rho$ for the Poisson-input case, ultimately producing the same vacuum probability as the coherent case at D.}
\label{fig:5}
\end{figure}
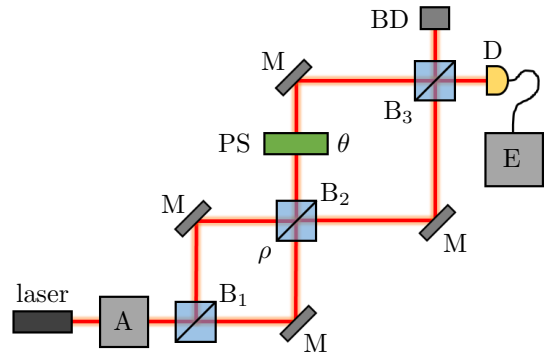

This is the most natural idea to suggest, since it ensures a stable phase relationship between the two coherent states after they are initially split from an input of \smash{$\rho_{\sqrt{2}\alpha}\otimes\rho_{0}$} to a preliminary 50:50 beam splitter prior to entering the VBS, as shown in \Fig{5}.
%_______________________________________________________________________________
%_______________________________________________________________________________

The coherent-generated input to the VBS is then $\rho=\rho_{\alpha}\otimes\rho_{-\alpha}$, yielding \smash{$p_{0}^{(1)}=e^{-|\alpha c_{\theta}-\alpha s_{\theta}|^{2}}=e^{-|\alpha|^{2}(1-s_{2\theta})}$}.

However, the case of initial Poisson input \smash{$\rho_{\sqrt{2}\alpha_{\overline \varphi}}\otimes\rho_{0}$} does \textit{not} produce a product of Poisson states after $\text{B}_1$, but instead produces a two-beam nondiagonal joint state $\rho$, given in (\ref{eq:B.1}) with $\alpha\to\sqrt{2}\alpha$.  As shown in \App{App.C}, also with $\alpha\to\sqrt{2}\alpha$, the action of the VBS on $\rho$ yields \smash{${\widetilde{\widetilde{p}}_{0}}\!\!\!\!{~}^{(1)}=e^{-|\alpha|^{2}(1-s_{2\theta})}$}, which is the same as the output for the coherent case, making this setup useless.

The trouble here is that the input to the VBS is not a product of Poisson states.
%                                End of II.C.1
%...............................................................................
%...............................................................................
%        II.C.2 Two Separate Identical Laser Beams (Succeeds but Fails)
\subsubsection{\label{sec:II.C.2}Two Separate Identical Laser Beams (Succeeds but Fails)}
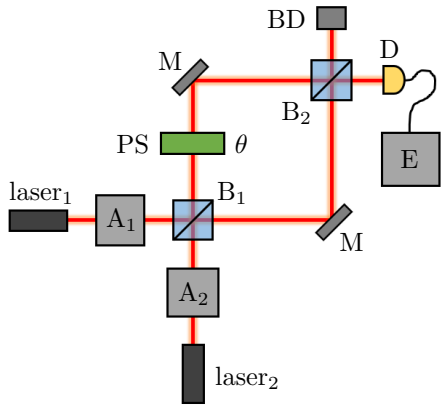
\begin{figure}[H]%                   FIGURE 6
\centering
\includegraphics[width=0.99\linewidth]{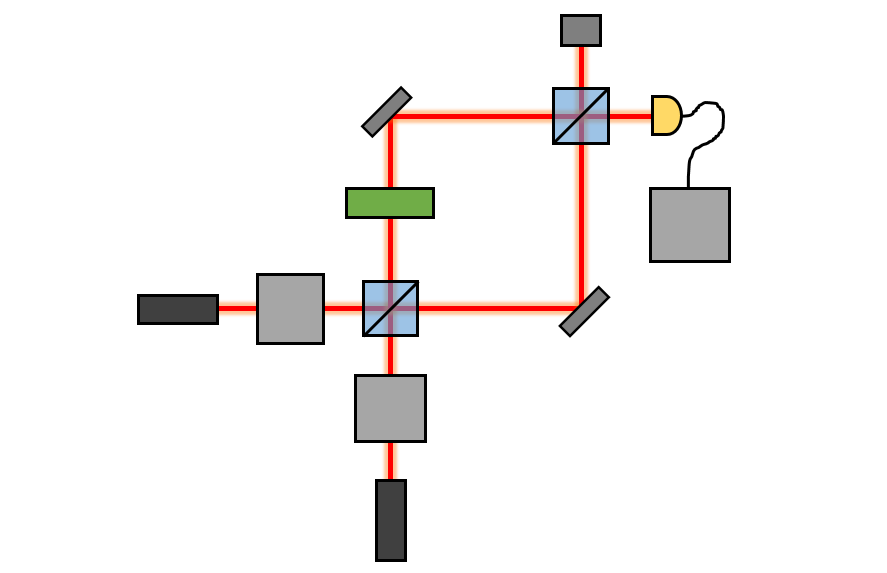}%Base name of figure file here.
\vspace{-12pt}
\setlength{\unitlength}{0.01\linewidth}
\begin{picture}(100,0)
\put(15.8,32){\normalsize $\text{laser}_1$}
\put(47.5,3.5){\normalsize $\text{laser}_2$}
\put(30.7,27.8){\normalsize $\text{A}_1$}
\put(41.8,16.4){\normalsize $\text{A}_2$}
\put(47.8,32){\normalsize $\text{B}_1$}
\put(38.7,52){\normalsize $\text{M}$}
\put(66.6,24.2){\normalsize $\text{M}$}
\put(32.3,39.3){\normalsize $\text{PS}$}
\put(50.5,39.2){\normalsize $\theta$}
\put(57.7,44.4){\normalsize $\text{B}_2$}
\put(55.5,58.5){\normalsize $\text{BD}$}%64.3,51.5
\put(76,37){\normalsize $\text{E}$}
\put(72.8,53.9){\normalsize $\text{D}$}
\end{picture}
\caption[]{(color online) Schematic of the two-laser realization to test for laser field state purity. All labels are the same as in \Fig{5}, except that here the VBS starts at $\text{B}_1$ and ends at $\text{B}_2$, and we now have two lasers and two attenuators.  For this test to work, it is essential that the two lasers be exactly the same model in every way, the mean wavelengths of each beam must be known, and the detection time-window must be very small, as discussed in the text.}
\label{fig:6}
\end{figure}

The idea here is to use two separate lasers of the same exact model and specifications as direct inputs to the VBS, as shown in \Fig{6}.  The input for the coherent case is then \smash{$\rho_{\alpha}\otimes\rho_{\alpha e^{i\chi}}$}, where we treat the phase of $\rho_{\alpha}$ as a reference so that $\text{arg}(\alpha)\equiv\varphi=0$, and $e^{i\chi}$ is the relative phase between the two beams.  The principal concern here is that $\chi$ will be arbitrarily varying over time, possibly rapidly.  Nevertheless, putting $\beta=\alpha e^{i\chi}$ in (\ref{eq:6}), the output of the VBS will lead to
%===============================================================================
\begin{equation}%                  Equation 9
p_{0}^{(1)}=e^{-|\alpha|^{2}(1+s_{2\theta}c_{\chi})},
\label{eq:9}
\end{equation}
%===============================================================================
which is the same as the quantity used to compute the coherent-generated curves in \Fig{4}.
%_______________________________________________________________________________
%_______________________________________________________________________________

In the case where both laser fields are Poisson states, their input to the VBS is the desired Poisson product \smash{$\rho_{\alpha_{\overline \varphi}}\otimes \rho_{\alpha_{\overline \chi}}$}, so that the VBS will produce 
%===============================================================================
\begin{equation}%                  Equation 10
\widetilde{p}_0^{(1)}  = e^{ - |\alpha |^2} I_0 (|\alpha |^{2}s_{2\theta } ).
\label{eq:10}
\end{equation}
%===============================================================================

While the output probabilities are now sufficiently different to distinguish the two cases of input states, what if relative phase $\chi$ varies randomly and rapidly?  In that case, only the coherent case is affected, and our measurements of \smash{$p_{0}^{(1)}$} will be an average over all relative classical phase shifts in (\ref{eq:9}), yielding 
%===============================================================================
\begin{equation}%                  Equation 11
p_{0_{\overline \chi } }^{(1)}  \equiv {\textstyle{1 \over {2\pi }}}\int_0^{2\pi } \!\!\!\!{p_0^{(1)} } d\chi  = e^{ - |\alpha |^2} I_0 (|\alpha |^{2}s_{2\theta } ),
\label{eq:11}
\end{equation}
%===============================================================================
which is \textit{identical} to the Poisson case, making this useless as well (even the general case of $\rho_{\alpha}\otimes\rho_{\beta e^{i\chi}}$ phase-averages to the general Poisson case of \smash{$e^{ - |\alpha c_\theta |^2  - |\beta s_\theta |^2 } I_0 (|\alpha ||\beta |s_{2\theta } )$}, where we used $\int\nolimits_{0}^{\pi}{e^{z\cos(x)}dx}=\pi I_{0}(z)$ from \cite[]{GraR}).

So in order for this setup to be useful, we need to find conditions for which the phase difference between the two lasers is relatively stable.  Unfortunately, there are two mechanisms working against us; the natural time-evolution of coherent states will cause rapid periodic changes in phase making it \textit{effectively} random, and the two different lasers are also subject to different random walks in their own phases.  Even if the random walks of the laser phases are negligible over the time-window, the time-evolution effects will still cause the output to be indistinguishable from that of Poisson inputs under presently achievable conditions.

To show that (\ref{eq:11}) is a valid model of what to expect, recall that the time-evolution of a coherent state in a harmonic oscillator potential is 
%===============================================================================
\begin{equation}%                  Equation 12
|\alpha (t)\rangle  = e^{ - i{\textstyle{\omega  \over 2}}t} |\alpha e^{ - i\omega t} \rangle,
\label{eq:12}
\end{equation}
%===============================================================================
where $\omega$ is the angular frequency of the field and $t$ is time.  Thus, supposing that the two lasers are attenuated to the same amplitude $|\alpha|$ but have slightly different frequencies $\omega_1$ and $\omega_2$, so that \smash{$\alpha_{1}\equiv|\alpha|e^{i(\varphi-\omega_{1}t)}$} and \smash{$\alpha_{2}\equiv|\alpha|e^{i(\phi-\omega_{2}t)}$}, then from (\ref{eq:6}), we again obtain (\ref{eq:9}), where now $\chi$ has time-dependence as
%===============================================================================
\begin{equation}%                  Equation 13
\chi\equiv \chi(t)=\chi_{0}-(\omega_{2}-\omega_{1})t,
\label{eq:13}
\end{equation}
%===============================================================================
where $\chi_{0}\equiv\phi-\varphi$ represents constant phase shifts due to differences in path length between the two beams, and we ignore effects due to the separate random walks of each laser phase.

Now, using the fact that \smash{$\omega _2  - \omega _1  = 2\pi c( {{\textstyle{1 \over {\lambda _2 }}} - {\textstyle{1 \over {\lambda _1 }}}})$}, and supposing that our two lasers only differ by a small fraction of nominal wavelength $\lambda$ such as ${\textstyle{1 \over {100}}}\lambda$, then if $\lambda_2 = 635[\text{nm}]$ and $\lambda_1 = 635.01[\text{nm}]$, we get $\omega _2  - \omega _1 \approx 4.67 \times 10^{10} [{\textstyle{{\text{rad}} \over \text{s}}}]$.  Thus, if the detection time-windows are typically on the order of $1[\mu\text{s}]$, then $\chi(t)$ will rotate with an angular speed of $4.67 \times 10^{4} [{\textstyle{{\text{rad}} \over {\mu \text{s}}}}]$, meaning that $\chi(t)$ cycles through a full $2\pi$ radians nearly $7500$ times per $1[\mu \text{s}]$-detection time-window.

Therefore, over the sample time, the relative phase $\chi(t)$ changes rapidly, and when a particular collapse happens at the detector, the phase at that moment is picked from a uniformly distributed set of angles over $[0,2\pi)$ because the phase $\chi(t)$ is linear in $t$.  Therefore, we are justified in supposing that even small differences in the laser wavelength will cause the measurements of the probability to be an average over the relative phase angle $\chi(t)$.

One possible exception is if we happen to get \smash{$\omega _2  = \omega _1$} exactly. Then if the source is a coherent state, $p$ will simply be one of the curves shown in \Fig{4}, and the test will work, indicating whether the random walks are negligible over the detection time-window.

Another possible way this method could work is if the lasers have tunable wavelengths with a fine-enough resolution to cause their relative phase shifts to be essentially constant over the detection time-window.  Unfortunately, present tuning techniques only have down to $0.1[\text{nm}]$ resolution, so it would be very hard to achieve this condition except by chance.

An interesting alternative would be to use \textit{chirping}, meaning the application of a high-speed phase shift by changing the optical path length over time in a way that cancels out the frequency-dependent evolution.  Then, if the chirping can be done reliably enough for each laser, it would not matter whether their frequencies were the same, since each state would have a constant net phase.

However, there is another way this could work.  If one has access to electronics capable of implementing detection time-windows on the order of $100[\text{fs}]$, then supposing that the lasers are nominally within ${\textstyle{1 \over {10}}}\lambda$, so that for instance, $\lambda_2 = 635[\text{nm}]$ and $\lambda_1 = 635.1[\text{nm}]$, then the resulting angular speed of $\chi(t)$ would be $4.67 \times 10^{-2} [{\textstyle{{\text{rad}} \over {100\text{fs}}}}]$, meaning that $\chi(t)$ only changes by $0.00744$ of a period, or about $2.7^{\circ}$ over the sample time.  Unfortunately, typical detectors have a dead time of about $100[\text{ns}]$, after which the phases would have changed so much that it would be hard to synchronize the next time window to fall at the same part of the period of $\chi(t)$.  Therefore, one would need a specialized segmented detector capable of taking a statistically significant number of consecutive measurements with zero effective dead time between each, or else a detector with dead time $\leq 100[\text{fs}]$.  Note that a \textit{statistically significant number of measurements} is the number of Bernoulli trials that lets the error in the estimators of the probabilities in the binomial experiment be acceptable for a given application.

Thus, (\ref{eq:9}) and (\ref{eq:10}) show that when the classical phase shift $\chi$ is relatively stable, the results of the setup in \Fig{6} will be identical to those in \Fig{3} (where $\chi=0$) and \Fig{4}.  However, since \Fig{6} uses two separate lasers as inputs, under presently achievable conditions, the relative phase $\chi$ is likely to cause the probability measurements to be an average over all values of $\chi$, yielding (\ref{eq:11}), so the coherent-input case would be indistinguishable from the Poisson case.  As explained above, technological advances in the time-resolution of detectors could eventually make this test a feasible means of distinguishing the cases of coherent vs.~Poisson input.  Otherwise, one must use quadrature methods, as in \cite[]{MDB}. Another method is that of \cite[]{BaWo}, but if limited to coherent or Poisson probes, that only works if the beam splitter is \textit{variable}, and one reaches the same conclusions as the present work.

At the very least, it is interesting to note that, from an instantaneous perspective, the coherent case actually has different vacuum probabilities than the Poisson case.  This is fascinating from a theoretical point of view because these kinds of highly simple on-off measurements are not usually enough to distinguish these two kinds of states.  Therefore the problems we encountered are only practical difficulties, and may eventually be overcome as the state of technology advances, making this method ultimately more useful in that case.
%                               End of II.C.2
%...............................................................................
%                                End of II.C
%-------------------------------------------------------------------------------
%                                END of II
%*******************************************************************************
%*******************************************************************************
%                             III. CONCLUSIONS
\section{\label{sec:IV}Conclusions}
The main result of this paper is that if high-enough time resolution can be achieved in the measurement process, then it \textit{is} possible to distinguish coherent states from phase-mixed coherent states (Poisson states) using only a variable beam splitter (VBS) and a single-photon on-off detector.  The reason this is possible is that the vacuum probability for one output beam of the VBS is generally different for the two cases where the input to the VBS is a product of coherent states or a product of Poisson states, as shown in \Sec{II.A}.  Therefore, the case of product-Poisson input was certainly worthwhile to study since it yields a simple test to verify whether a particular laser is truly in a pure coherent state over a given detection time-window.

Then in \Sec{II.B}, we took a closer look at phase shifts in the coherent-state input case, and verified that the resulting detection probabilities are reliably distinguishable from those of the Poisson-input case, provided that that the mean photon numbers of each beam are not too close to zero, and not so large than they cause their respective beams' detection probabilities to be close to $1$.  It was found that choosing a nonvacuum detection probability for each independent beam of approximately $p_{*}=0.85$ produces a favorable condition for getting useful results.  This section implicitly assumed that the phases of each input beam in the coherent case were constant with respect to each other, but later in \Sec{II.C}, we investigated the case of time-dependent phase.

\Section{II.C} investigated two simple proposals to realize a test to distinguish coherent states from Poisson states based on the results of \Sec{II.A} and \Sec{II.B}.  The first method in \Sec{II.C.1} uses a preliminary beam splitter and mirrors to prepare inputs to the VBS that have a stable phase relationship in the coherent-input case since they come from a single beam.  However, it was shown in \App{App.B} and \App{App.C} that the vacuum probabilities at the detector would be identical for both cases of coherent input and Poisson input.  The problem was seen to arise from the fact that the preliminary 50:50 beam splitter's action on a Poisson state does \text{not} cause a product of Poisson states, but rather it causes a two-beam nondiagonal joint state, as shown in (\ref{eq:B.1}).  Therefore, the setup shown in \Fig{5} is useless for the desired test.

An alternative was then given in \Sec{II.C.2}, which considered a two-laser input to the VBS as shown in \Fig{6}.  This test requires that the two lasers be the exact same model, and allows the product of Poisson inputs to be realizable if the lasers' phase diffusion occurs much more rapidly than the given time-window.  However, the use of two separate lasers will result in a slight difference in wavelength of their fields due to unavoidable differences in laser construction, temperature, etc.  Thus, under typical conditions, the natural time-evolution of coherent states would cause the relative phase between the two input fields to vary rapidly over the detection time-window.  The instantaneous output probability is then a function of time, due to its dependence on relative phase, and since this phase is a uniformly distributed random variable, the estimator of the probability will be a time average of the instantaneous function, resulting in probabilities that make the two cases of coherent input and Poisson input indistinguishable, despite their instantaneous differences.

However, \Sec{II.C.2} then considered the conditions for which a typical laser could be tested that would avoid the problems of effective phase averaging due to time-evolution, and found that if the data acquisition electronics are capable of time-windows on the order of $100[\text{fs}]$, and if the detector has dead times (or effective dead times if partitioned) on this timescale as well, then it would be possible to observe the differences between the two kinds of input states using this method.  Since this time-resolution is not currently widely available and may not be presently possible, this method must only be considered a theoretical curiosity unless further advances are made in electronics.  However, since it was not previously thought that on-off statistics could distinguish coherent states from Poisson states, then this at least advances our understanding of what is theoretically possible.

Throughout this treatment, detectors were treated as ideal, but a real detector can be modeled as an ideal detector preceded by a VBS with transmittivity $T=\eta$, where $\eta$ is the quantum efficiency of the detector.  For an even more realistic model, the auxiliary input to this VBS (perpendicular to the detector) can be thought of as a thermal state with its mean photon number chosen so that it explains the dark counts measured by the detector, found by setting the primary input to vacuum and measuring the counts.  However, in most cases, single-photon detectors are thermo-electrically cooled to the point where the dark counts are negligible, meaning that a pure state at the primary input will be \textit{almost perfectly} transmitted to the imaginary ideal detector.  Specifically, the mixture induced by the thermal state modeling the dark noise will be a mixed state where the pure state with the largest probability is the primary input state, and its probability will be very near $1$.

Certainly there are other ways to see if a laser's field is in a coherent state, and many others \textit{have} done this.  However, the goal here was to design a \textit{simple} test with a minimum of hardware.  The proposed setup in \Fig{6} fits the bill nicely, with the only drawbacks being that two identical lasers are needed which then requires ultrafast electronics to resolve the changes due to the time-dependent relative phase.  Nevertheless, as detection devices improve, this test may become feasible.  

Since this paper is a theoretical treatment only, it would be interesting to see an experimental implementation of this test, though it is likely to be difficult to achieve the time-resolution needed.  Hopefully, when electronics have improved to the time-sensitivity required, this proposed test will serve as a relatively quick and inexpensive method for verifying that a particular laser device is adequate as a source of coherent (or Poisson) states for a given application.
%                                END of III
%*******************************************************************************
%*******************************************************************************
%                               ACKNOWLEDGEMENTS
\begin{acknowledgments}
This work was partially funded by the Innovation and Entrepreneurship Doctoral Fellowship at Stevens Institute of Technology.  Special thanks to Ting Yu at Stevens I.T. and Joan Hoffmann at The Johns Hopkins University Applied Physics Laboratory for helpful feedback.
\end{acknowledgments}
%                            END of ACKNOWLEDGEMENTS
%*******************************************************************************
%*******************************************************************************
%                                   APPENDIX
\begin{appendix}
%-------------------------------------------------------------------------------
%    App.A. Vacuum Probability of One Output of a Variable Beam Splitter with Two Poisson Inputs
\section{\label{sec:App.A}Vacuum Probability of One Output of a Variable Beam Splitter with Two Poisson Inputs}
Here we will derive (\ref{eq:8}), which is a crucial result of this paper.  Recalling that \smash{$\widetilde{p}_0^{(1)}  \equiv \langle 0|\text{tr}_2 (\widetilde{\rho}  )|0\rangle$} where $\widetilde{\rho}\equiv B_\theta  (\rho _{\alpha _{\overline \varphi } }  \otimes \rho _{\beta _{\overline \phi } } )B_\theta  ^{\dag}$ is given in (\ref{eq:7}), we obtain
%===============================================================================
\begin{equation}%                  Equation A.1
\widetilde{p}_0^{(1)}\!  =\!  {\left(\!\! {\begin{array}{*{20}l}
   {e^{ - |\alpha |^2  - |\beta |^2 }\sum\limits_{k = 0}^\infty{} \sum\limits_{n = 0}^\infty  {\sum\limits_{m = 0}^\infty  {\!\left( {{\textstyle{{(|\alpha |c_\theta )^{n} } \over {n!}}}} \right)^{2}\!\! \left( {{\textstyle{{(|\beta |s_\theta )^{m} } \over {m!}}}} \right)^{2} } } }  \\
   { \times\! \sum\limits_{q = 0}^n {\sum\limits_{r = 0}^m {\!\binom{n}{q}\binom{m}{r}\!\!\left(\! {{\textstyle{{ - s_\theta  } \over {c_\theta  }}}}\! \right)^q \!\!\left(\! {{\textstyle{{c_\theta  } \over {s_\theta  }}}}\! \right)^r \!\!\delta _{q + r,k} \delta _{n + m,q + r} } } }  \\
   { \times\! \sum\limits_{s = 0}^n {\sum\limits_{t = 0}^m {\!\binom{n}{s}\binom{m}{t}\!\!\left(\! {{\textstyle{{ - s_\theta  } \over {c_\theta  }}}}\! \right)^s \!\!\left(\! {{\textstyle{{c_\theta  } \over {s_\theta  }}}}\! \right)^t \!\!\delta _{s + t,k} \delta _{n + m,s + t} } } }  \\
   { \times \sqrt {\!(n\! +\! m\! -\! q\! -\! r)!(q\! +\! r)!(n\! +\! m\! -\! s\! -\! t)!(s\! +\! t)!} }  \\
\end{array}}\!\! \right)\!,\!}
\label{eq:A.1}
\end{equation}
%===============================================================================
where \smash{$\binom{n}{m}\equiv\frac{n!}{m!(n-m)!}$}, and $\delta_{n,m}$ is the Kronecker delta.  The deltas cause $q+r=n+m=k$ and $s+t=n+m=k$, from which we can make several conclusions.  First, the fact that $n+m=k$ constrains those indices so that $\text{max}(n)=k$ and $\text{max}(m)=k$, so we can replace the upper limits of the $n$ and $m$ sums with $k$, still subject  to $n+m=k$.  Similar truncations happen for the other indices, but for now, we can simply rewrite (\ref{eq:A.1}) as
%===============================================================================
\begin{equation}%                  Equation A.2
\widetilde{p}_0^{(1)}\!  =\!  {\left(\!\! {\begin{array}{*{20}l}
   {e^{ - |\alpha |^2  - |\beta |^2 }\sum\limits_{k = 0}^\infty{} \sum\limits_{n = 0}^k  {\sum\limits_{m = 0}^k  {\!\left( {{\textstyle{{(|\alpha |c_\theta )^{n} } \over {n!}}}} \right)^{2}\!\! \left( {{\textstyle{{(|\beta |s_\theta )^{m} } \over {m!}}}} \right)^{2} } } }  \\
   { \times\! \sum\limits_{q = 0}^n {\sum\limits_{r = 0}^m {\!\binom{n}{q}\binom{m}{r}\!\!\left(\! {{\textstyle{{ - s_\theta  } \over {c_\theta  }}}}\! \right)^q \!\!\left(\! {{\textstyle{{c_\theta  } \over {s_\theta  }}}}\! \right)^r } } }  \\
   { \times\! \sum\limits_{s = 0}^n {\sum\limits_{t = 0}^m {\!\binom{n}{s}\binom{m}{t}\!\!\left(\! {{\textstyle{{ - s_\theta  } \over {c_\theta  }}}}\! \right)^s \!\!\left(\! {{\textstyle{{c_\theta  } \over {s_\theta  }}}}\! \right)^t } }\sqrt {0!k!0!k!} }  \\
\end{array}}\!\! \right)\!,\!}
\label{eq:A.2}
\end{equation}
%===============================================================================
subject to $m=k-n$, $r=k-q$, and $t=k-s$.  For the $m$ sum, the fact that the original sum extended to $\infty$ means that the $m=k-n\leq\infty$ will simply pick that term out of the sum.  

However, the situation is slightly different for the other sums.  For the $r$ sum, we have $r\leq m$, which becomes $r\leq k-n$.  But then, applying $r=k-q$, we get $k-q\leq k-n$, which leads to $q\geq n$, and since the $q$ sum has an inherent upper limit of $n$, then this means that only the equality case of $q=n$ survives, so that these constraints effectively act as $\delta_{q,n}\delta_{r,k-n}$, where the second delta is due to the constraint $r=k-q$ with $q=n$ applied to it.  Similarly, the constraints on $s$ and $t$ effectively insert $\delta_{s,n}\delta_{t,k-n}$, all of which simplifies (\ref{eq:A.2}) to
%===============================================================================
\begin{equation}%                  Equation A.3
\widetilde{p}_0^{(1)}\!  =\!  {\left(\!\! {\begin{array}{*{20}l}
   {e^{ - |\alpha |^2  - |\beta |^2 }\sum\limits_{k = 0}^\infty{} \sum\limits_{n = 0}^k  {  {\!\left( {{\textstyle{{(|\alpha |c_\theta )^{n} } \over {n!}}}} \right)^{2}\!\! \left( {{\textstyle{{(|\beta |s_\theta )^{k-n} } \over {(k-n)!}}}} \right)^{2} } } }  \\
   { \times\! \left({ {\!\binom{n}{n}\binom{k-n}{k-n}\!\!\left(\! {{\textstyle{{ - s_\theta  } \over {c_\theta  }}}}\! \right)^n \!\!\left(\! {{\textstyle{{c_\theta  } \over {s_\theta  }}}}\! \right)^{k-n} } }\right)^{2} k! }  \\
\end{array}}\!\! \right)\!,\!}
\label{eq:A.3}
\end{equation}
%===============================================================================
which is then easy to rewrite as
%===============================================================================
\begin{equation}%                  Equation A.4
\widetilde{p}_0^{(1)}  = e^{ - |\alpha |^2  - |\beta |^2 } \sum\limits_{k = 0}^\infty  {\textstyle{1 \over {k!}}}{\sum\limits_{n = 0}^k {\left[ {{\textstyle \binom{k}{n}}(|\alpha |s_\theta  )^n (|\beta |c_\theta  )^{k - n} } \right]^{2}\!\!\!, } }
\label{eq:A.4}
\end{equation}
%===============================================================================
which, abbreviating $A \equiv |\alpha |s_\theta$ and $B \equiv |\beta |c_\theta$, becomes
%===============================================================================
\begin{equation}%                  Equation A.5
\widetilde{p}_0^{(1)}  = e^{ - |\alpha |^2  - |\beta |^2 } \sum\limits_{k = 0}^\infty  {\textstyle{1 \over {k!}}}{\sum\limits_{n = 0}^k {\left[ {\textstyle {\binom{k}{n}}A^n B^{k - n} } \right]^2 } }\!.
\label{eq:A.5}
\end{equation}
%===============================================================================

Next, we compute terms contributing to the total vacuum probability \smash{$\widetilde{p}_0^{(1)}$} for fixed values of index $k$, rescaling as \smash{$p_k ' \equiv e^{|\alpha |^2  + |\beta |^2 } (\widetilde{p}_0^{(1)} )_k$} to reduce clutter.  The first nine terms give
%===============================================================================
\begin{equation}%                  Equation A.6
\begin{array}{*{20}l}
   {p_0 '} &\!\! { = 1}  \\
   {p_1 '} &\!\! { = A^2  + B^2 }  \\
   {p_2 '} &\!\! { = {\textstyle{1 \over 2}}(A^4  + B^4 ) + 2A^2 B^2 }  \\
   {p_3 '} &\!\! { = {\textstyle{1 \over 6}}(A^6  + B^6 ) + {\textstyle{3 \over 2}}A^2 B^2 (A^2  + B^2 )}  \\
   {p_4 '} &\!\! { = {\textstyle{1 \over {24}}}(A^8  + B^8 ) + {\textstyle{2 \over 3}}A^2 B^2 (A^4  + B^4 ) + {\textstyle{3 \over 2}}A^4 B^4 }  \\
   {p_5 '} &\!\! { = \left( \begin{array}{l}
 {\textstyle{1 \over {120}}}(A^{10}  + B^{10} ) + {\textstyle{5 \over {24}}}A^2 B^2 (A^6  + B^6 ) \\ 
  + {\textstyle{5 \over 6}}A^4 B^4 (A^2  + B^2 ) \\ 
 \end{array} \right)}  \\
   {p_6 '} &\!\! { = \left( \begin{array}{l}
 {\textstyle{1 \over {720}}}(A^{12}  + B^{12} ) + {\textstyle{1 \over {20}}}A^2 B^2 (A^8  + B^8 ) \\ 
  + {\textstyle{5 \over {16}}}A^4 B^4 (A^4  + B^4 ) + {\textstyle{5 \over 9}}A^6 B^6  \\ 
 \end{array} \right)}  \\
   {p_7 '} &\!\! { = \left( \begin{array}{l}
 {\textstyle{1 \over {5040}}}(A^{14}  + B^{14} ) + {\textstyle{7 \over {720}}}A^2 B^2 (A^{10}  + B^{10} ) \\ 
  + {\textstyle{7 \over {80}}}A^4 B^4 (A^6  + B^6 ) + {\textstyle{{35} \over {144}}}A^6 B^6 (A^2  + B^2 ) \\ 
 \end{array} \right)}  \\
   {p_8 '} &\!\! { = \left(\! \begin{array}{l}
 {\textstyle{1 \over {40320}}}(A^{16}  + B^{16} ) + {\textstyle{1 \over {630}}}A^2 B^2 (A^{12}  + B^{12} ) \\ 
  + {\textstyle{7 \over {360}}}A^4 B^4 (A^8  + B^8 ) + {\textstyle{7 \over {90}}}A^6 B^6 (A^4  + B^4 ) \\ 
  + {\textstyle{{35} \over {288}}}A^8 B^8  \\ 
 \end{array}\! \right),}  \\
\end{array}
\label{eq:A.6}
\end{equation}
%===============================================================================
where again, $A \equiv |\alpha |s_\theta$ and $B \equiv |\beta |c_\theta$.

Our next step is to find a form with more predictable coefficients so we can write a general formula for each of these $k$-value terms.  To accomplish this, notice that we can use $(A^2  + B^2 )^k$ to derive an alternate form for each $p_k '$ in (\ref{eq:A.6}).  For example, for $k=2$,
%===============================================================================
\begin{equation}%                  Equation A.7
\begin{array}{*{20}l}
   {A^4  + 2A^2 B^2  + B^4 } &\!\! { = (A^2  + B^2 )^2 }  \\
   {{\textstyle{1 \over 2}}(A^4  + B^4 ) + A^2 B^2 } &\!\! { = {\textstyle{1 \over 2}}(A^2  + B^2 )^2 }  \\
   {{\textstyle{1 \over 2}}(A^4  + B^4 ) + 2A^2 B^2 } &\!\! { = {\textstyle{1 \over 2}}(A^2  + B^2 )^2  + A^2 B^2 .}  \\
\end{array}
\label{eq:A.7}
\end{equation}
%===============================================================================
Continuing for the other $k$-values shown, and further abbreviating with $Q \equiv A^2  + B^2$ and $R \equiv AB$, we get 
%===============================================================================
\begin{equation}%                  Equation A.8
\begin{array}{*{20}l}
   {p_0 '} &\!\! { = 1}  \\
   {p_1 '} &\!\! { = Q}  \\
   {p_2 '} &\!\! { = {\textstyle{1 \over 2}}Q^2  + R^2 }  \\
   {p_3 '} &\!\! { = {\textstyle{1 \over 6}}Q^3  + R^2 Q}  \\
   {p_4 '} &\!\! { = {\textstyle{1 \over {24}}}Q^4  + {\textstyle{1 \over 2}}R^2 Q^2  + {\textstyle{1 \over 4}}R^4 }  \\
   {p_5 '} &\!\! { = {\textstyle{1 \over {120}}}Q^5  + {\textstyle{1 \over 6}}R^2 Q^3  + {\textstyle{1 \over 4}}R^4 Q}  \\
   {p_6 '} &\!\! { = {\textstyle{1 \over {720}}}Q^6  + {\textstyle{1 \over {24}}}R^2 Q^4  + {\textstyle{1 \over 8}}R^4 Q^2  + {\textstyle{1 \over {36}}}R^6 }  \\
   {p_7 '} &\!\! { = {\textstyle{1 \over {5040}}}Q^7  + {\textstyle{1 \over {120}}}R^2 Q^5  + {\textstyle{1 \over {24}}}R^4 Q^3  + {\textstyle{1 \over {36}}}R^6 Q}  \\
   {p_8 '} &\!\! { = {\textstyle{1 \over {40320}}}Q^8 \! + {\textstyle{1 \over {720}}}R^2 Q^6 \! + {\textstyle{1 \over {96}}}R^4 Q^4 \! + {\textstyle{1 \over {72}}}R^6 Q^2 \! +\! {\textstyle{1 \over {576}}}R^{8}\!,}  \\
\end{array}
\label{eq:A.8}
\end{equation}
%===============================================================================
where it was necessary to reuse various expansions along the way.

Now, from (\ref{eq:A.8}), it is much easier to see a pattern and we can write the formula for the $k$th term as
%===============================================================================
\begin{equation}%                  Equation A.9
p_k '  = \sum\limits_{m = 0}^{M_k } {\left( {{\textstyle{{R^m } \over {m!}}}} \right)^2 \! {\textstyle{{Q^{k - 2m} } \over {(k - 2m)!}}}},
\label{eq:A.9}
\end{equation}
%===============================================================================
where \smash{$M_k  \equiv (k + {\textstyle{{( - 1)^k  - 1} \over 2}})/2$}, which is $k/2$ when $k$ is even, and is $(k-1)/2$ when $k$ is odd.

Now, expanding the first nine of these terms without simplifying, we get
%===============================================================================
\begin{equation}%                  Equation A.10
\begin{array}{*{20}l}
   {p_0 '} &\!\! { = (\!{\textstyle{{R^0 } \over {0!}}}\!)^2 {\textstyle{{Q^0 } \over {0!}}}}  \\
   {p_1 '} &\!\! { = (\!{\textstyle{{R^0 } \over {0!}}}\!)^2 {\textstyle{{Q^1 } \over {1!}}}}  \\
   {p_2 '} &\!\! { = (\!{\textstyle{{R^0 } \over {0!}}}\!)^2 {\textstyle{{Q^2 } \over {2!}}} \!+\! (\!{\textstyle{{R^1 } \over {1!}}}\!)^2 {\textstyle{{Q^0 } \over {0!}}}}  \\
   {p_3 '} &\!\! { = (\!{\textstyle{{R^0 } \over {0!}}}\!)^2 {\textstyle{{Q^3 } \over {3!}}} \!+\! (\!{\textstyle{{R^1 } \over {1!}}}\!)^2 {\textstyle{{Q^1 } \over {1!}}}}  \\
   {p_4 '} &\!\! { = (\!{\textstyle{{R^0 } \over {0!}}}\!)^2 {\textstyle{{Q^4 } \over {4!}}} \!+\! (\!{\textstyle{{R^1 } \over {1!}}}\!)^2 {\textstyle{{Q^2 } \over {2!}}} \!+\! (\!{\textstyle{{R^2 } \over {2!}}}\!)^2 {\textstyle{{Q^0 } \over {0!}}}}  \\
   {p_5 '} &\!\! { = (\!{\textstyle{{R^0 } \over {0!}}}\!)^2 {\textstyle{{Q^5 } \over {5!}}} \!+\! (\!{\textstyle{{R^1 } \over {1!}}}\!)^2 {\textstyle{{Q^3 } \over {3!}}} \!+\! (\!{\textstyle{{R^2 } \over {2!}}}\!)^2 {\textstyle{{Q^1 } \over {1!}}}}  \\
   {p_6 '} &\!\! { = (\!{\textstyle{{R^0 } \over {0!}}}\!)^2 {\textstyle{{Q^6 } \over {6!}}} \!+\! (\!{\textstyle{{R^1 } \over {1!}}}\!)^2 {\textstyle{{Q^4 } \over {4!}}} \!+\! (\!{\textstyle{{R^2 } \over {2!}}}\!)^2 {\textstyle{{Q^2 } \over {2!}}} \!+\! (\!{\textstyle{{R^3 } \over {3!}}}\!)^2 {\textstyle{{Q^0 } \over {0!}}}}  \\
   {p_7 '} &\!\! { = (\!{\textstyle{{R^0 } \over {0!}}}\!)^2 {\textstyle{{Q^7 } \over {7!}}} \!+\! (\!{\textstyle{{R^1 } \over {1!}}}\!)^2 {\textstyle{{Q^5 } \over {5!}}} \!+\! (\!{\textstyle{{R^2 } \over {2!}}}\!)^2 {\textstyle{{Q^3 } \over {3!}}} \!+\! (\!{\textstyle{{R^3 } \over {3!}}}\!)^2 {\textstyle{{Q^1 } \over {1!}}}}  \\
   {p_8 '} &\!\! { = (\!{\textstyle{{R^0 } \over {0!}}}\!)^2 {\textstyle{{Q^8 } \over {8!}}} \!+\! (\!{\textstyle{{R^1 } \over {1!}}}\!)^2 {\textstyle{{Q^6 } \over {6!}}} \!+\! (\!{\textstyle{{R^2 } \over {2!}}}\!)^2 {\textstyle{{Q^4 } \over {4!}}} \!+\! (\!{\textstyle{{R^3 } \over {3!}}}\!)^2 {\textstyle{{Q^2 } \over {2!}}} \!+\! (\!{\textstyle{{R^4 } \over {4!}}}\!)^2 {\textstyle{{Q^0 } \over {0!}}}}.  \\
\end{array}
\label{eq:A.10}
\end{equation}
%===============================================================================
Now, since all of the $p_{k}'$ must be summed to get the full vacuum probability, we are free to group them any way that is convenient.  Thus, after doing the sum over all $p_{k}'$, if we group all terms containing particular $m$-values of \smash{$( {{\textstyle{{R^m } \over {m!}}}})^2$}, we notice that each group contains the same exponential series,
%===============================================================================
\begin{equation}%                  Equation A.11
\begin{array}{*{20}l}
   {\widetilde{p}_0^{(1)} } &\!\! { = e^{ - |\alpha |^2  - |\beta |^2 } \!\sum\limits_{k = 0}^\infty  {p_k '} }  \\
   {} &\!\! { = e^{ - |\alpha |^2  - |\beta |^2 }\!\! \left(\! {(\!{\textstyle{{R^0 } \over {0!}}}\!)^2 \!\left({\sum\limits_{t = 0}^\infty  {\!{\textstyle{{Q^t } \over {t!}}}}}\!\right) \! +\! (\!{\textstyle{{R^1 } \over {1!}}}\!)^2 \!\left({\sum\limits_{u = 0}^\infty  {\!{\textstyle{{Q^u } \over {u!}}}}}\!\right)\!  +  \cdots }\! \right)}  \\
   {} &\!\! { = e^{ - |\alpha |^2  - |\beta |^2 } e^Q \sum\limits_{k = 0}^\infty  {({\textstyle{{R^k } \over {k!}}})^2 } }.  \\
\end{array}
\label{eq:A.11}
\end{equation}
%===============================================================================
Then, using the facts that $Q = |\alpha |^2 s_\theta ^2  + |\beta |^2 c_\theta ^2$ and $R = |\alpha ||\beta |s_\theta  c_\theta   = {\textstyle{1 \over 2}}|\alpha ||\beta |s_{2\theta }$, (\ref{eq:A.11}) becomes
%===============================================================================
\begin{equation}%                  Equation A.12
\begin{array}{*{20}l}
   {\widetilde{p}_0^{(1)} } &\!\! { = e^{ - |\alpha |^2  - |\beta |^2  + |\alpha |^2 s_\theta ^2  + |\beta |^2 c_\theta ^2 } \sum\limits_{k = 0}^\infty  {\left( {{\textstyle{{\left({{\textstyle{1 \over 2}}|\alpha ||\beta |s_{2\theta }}\right)^k } \over {k!}}}} \right)^2 } }  \\
   {} &\!\! { = e^{ - |\alpha |^2 (1 - s_\theta ^2 ) - |\beta |^2 (1 - c_\theta ^2 )} \sum\limits_{k = 0}^\infty  {{\textstyle{{\left( {{\textstyle{1 \over 4}}(|\alpha ||\beta |s_{2\theta } )^2 } \right)^k } \over {(k!)^2 }}}} }  \\
   {} &\!\! { = e^{ - |\alpha c_\theta |^2  - |\beta s_\theta |^2  } I_0 (|\alpha ||\beta |s_{2\theta } )},  \\
\end{array}
\label{eq:A.12}
\end{equation}
%===============================================================================
which is the result in (\ref{eq:8}), where again, $I_0 (x)$ is the modified Bessel function of the first kind, order zero.

Furthermore, by comparing (\ref{eq:A.5}) with (\ref{eq:A.12}) and also (\ref{eq:A.9}), we get the following interesting identities;
%===============================================================================
\begin{equation}%                  Equation A.13
\begin{array}{*{20}l}
   {\sum\limits_{n = 0}^\infty  {{\textstyle{1 \over {n!}}}\sum\limits_{k = 0}^n {\left[ {\binom{n}{k}A^{n - k} B^k } \right]^2 } } } &\!\! { = e^{A^2  + B^2 } I_0 (AB)}  \\
   {\sum\limits_{n = 0}^\infty  {n!\sum\limits_{k = 0}^n {\left[ {{\textstyle{{A^{n - k} } \over {(n - k)!}}}{\textstyle{{B^k } \over {k!}}}} \right]^2 } } } &\!\! { = e^{A^2  + B^2 } I_0 (AB)}  \\
   {\sum\limits_{n = 0}^\infty  {\sum\limits_{k = 0}^{K_n } {\left( {{\textstyle{{(AB)^k } \over {k!}}}} \right)^2 {\textstyle{{(A^2  + B^2 )^{n - 2k} } \over {(n - 2k)!}}}} } } &\!\! { = e^{A^2  + B^2 } I_0 (AB),}  \\
\end{array}
\label{eq:A.13}
\end{equation}
%===============================================================================
where \smash{$K_n \equiv (n + {\textstyle{{( - 1)^n  - 1} \over 2}})/2$}, $A$ and $B$ are real, and we have relabeled various quantities to give them a more standard appearance.  Also, due to the right sides of (\ref{eq:A.13}) all being the same, all the left sides are equal to each other, as well.
%                                 End App.A
%-------------------------------------------------------------------------------
%-------------------------------------------------------------------------------
%    App.B. Outputs of a 50:50 Beam Splitter with Poisson State and Vacuum Inputs
\section{\label{sec:App.B}Outputs of a 50:50 Beam Splitter with Poisson State and Vacuum Inputs}
Here, we will briefly show that when a Poisson state enters a 50:50 beam splitter with the vacuum state $|0\rangle$ at the other input port, the joint-beam state is \textit{not} a product of Poisson states, but if one measures one of the beams, then the other will be in a Poisson state.

First, since (\ref{eq:7}) is the output of a VBS given two Poisson inputs, then here, we simply set $\beta=0$ to get vacuum on one of the inputs, and set $\theta=45^{\circ}$ to get 50:50 behavior from the VBS.  Then, $0^m$ acts as $\delta _{m,0}$ since only $0^0 =1$ in that sum, so $m=0$, which also causes \smash{$\delta _{r,0}$} and \smash{$\delta _{t,0}$}, leading to total state \smash{$\rho\equiv B_{45^{\circ}}(\rho_{\alpha_{\overline \varphi}}\otimes\rho_0 )B_{45^{\circ}}^{\dag}$},
%===============================================================================
\begin{equation}%                  Equation B.1
\begin{array}{*{20}l}
   {\rho  = } &\!\! {e^{ - |\alpha |^2 } \sum\limits_{n = 0}^\infty  {\left( {{\textstyle{{\left({|\alpha |/\sqrt 2 }\right)^n } \over {n!}}}} \right)^2 \sum\limits_{q = 0}^n {{\textstyle{{n!( - 1)^q } \over {\sqrt {(n - q)!q!} }}}\sum\limits_{s = 0}^n {{\textstyle{{n!( - 1)^s } \over {\sqrt {(n - s)!s!} }}}} } } }  \\
   {} &\!\! { \times |n - q\rangle \langle n - s| \otimes |q\rangle \langle s|.}  \\
\end{array}
\label{eq:B.1}
\end{equation}
%===============================================================================
Notice that (\ref{eq:B.1}) is not diagonal in each subsystem, and therefore cannot be a product of Poisson states, which would have to be diagonal.

To get the state of a single output beam, apply the partial trace over subsystem 2, which introduces $\delta _{q,k}$ and $\delta _{s,k}$ which can be handled by first applying the effective $\delta _{q,s}$ and then applying $\delta _{q,k}$, which results in
%===============================================================================
\begin{equation}%                  Equation B.2
\rho ^{(1)}  = e^{ - |\alpha |^2 } \sum\limits_{n = 0}^\infty  {\left( {{\textstyle{{\left({|\alpha |/\sqrt 2}\right)^n } \over {n!}}}} \right)^2 n!\sum\limits_{k = 0}^n {{\textstyle\binom{n}{k}}|n - k\rangle \langle n - k|.} }
\label{eq:B.2}
\end{equation}
%===============================================================================
Writing out terms for fixed $n$ reveals the alternative form,
%===============================================================================
\begin{equation}%                  Equation B.3
\rho ^{(1)}  = e^{ - |\alpha |^2 } \sum\limits_{m = 0}^\infty  {\sum\limits_{n = m}^\infty  {{\textstyle{{\left({|\alpha |^2 /2}\right)^n } \over {n!}}}{\textstyle\binom{n}{m}}} |m\rangle \langle m|,}
\label{eq:B.3}
\end{equation}
%===============================================================================
which can then be simplified as
%===============================================================================
\begin{equation}%                  Equation B.4
\begin{array}{*{20}l}
   {\rho ^{(1)} } &\!\! { = e^{ - |\alpha |^2 } \sum\limits_{m = 0}^\infty  {{\textstyle{{\left( {|\alpha |^2 /2} \right)^m } \over {m!}}}|m\rangle \langle m|\sum\limits_{n - m = 0}^\infty \!\! {{\textstyle{{\left( {|\alpha |^2 /2} \right)^{n - m} } \over {(n - m)!}}}} } }  \\
   {} &\!\! { = e^{ - |\alpha |^2 } e^{{\textstyle{{|\alpha |^2 } \over 2}}} \sum\limits_{m = 0}^\infty  {{\textstyle{{\left( {|\alpha |^2 /2} \right)^m } \over {m!}}}|m\rangle \langle m|} }  \\
   {} &\!\! { = e^{ - |{\textstyle{\alpha  \over {\sqrt 2 }}}|^2 } \sum\limits_{m = 0}^\infty  {{\textstyle{{\left( {|{\textstyle{\alpha  \over {\sqrt 2 }}}|^2 } \right)^m } \over {m!}}}|m\rangle \langle m|} }  \\
   {} &\!\! { = \rho _{{\textstyle{\alpha  \over {\sqrt 2 }}}_{\overline \varphi } }, }  \\
\end{array}
\label{eq:B.4}
\end{equation}
%===============================================================================
where our notation in the last line comes from (\ref{eq:3}), and means that this beam is in a Poisson state of mean photon number $|{\textstyle{\alpha  \over {\sqrt 2 }}}|^2$ \textit{only if the other beam is continually measured}.  Partial-tracing to get the state of the other beam yields the same result.

The important thing to note here is that one cannot make a product of Poisson states by applying a 50:50 beam splitter to a Poisson state in product with the vacuum.  This means that although the vacuum probability of the output of a VBS with product-Poisson inputs is different from that caused by product-coherent inputs, a fully phase-damped laser field does not yield a product of Poisson states after a 50:50 beam splitter.  That is the motivation for the modified test suggested in \Fig{6}.
%                                  End App.B
%-------------------------------------------------------------------------------
%-------------------------------------------------------------------------------
%    App.C. Single-Beam Vacuum Probability of VBS with Joint Beam Input Made from 50:50 Split of Poisson and Vacuum
\section{\label{sec:App.C}Single-Beam Vacuum Probability of VBS with Joint Beam Input Made from 50:50 Split of Poisson and Vacuum}
Here, we will derive the output we should expect from a VBS when the input is a joint-beam state $\rho$ prepared by sending a Poisson state in product with the vacuum $\rho_{\alpha_{\overline \varphi}}\otimes\rho_0$ into a 50:50 beam splitter (BS).  The input to the VBS is $\rho$ as shown just prior to $\text{B}_2$ in \Fig{5}, and was given in (\ref{eq:B.1}).

Our goal now is to find the joint-beam output state of the VBS when its input is $\rho$ from (\ref{eq:B.1}).  Thus, computing \smash{$\widetilde{\widetilde{\rho}}\equiv B_{\theta}\rho B_{\theta}^{\dag}$}, and using the binomial theorem to handle the action of $B_{\theta}$ on the creation and annihilation operators, after some reorganization, we arrive at the joint-beam output state,
%===============================================================================
\begin{equation}%                  Equation C.1
\begin{array}{*{20}l}
   {\widetilde{\widetilde{\rho}}  = } &\!\! {e^{ - |\alpha |^2 } \sum\limits_{n = 0}^\infty  {\left( {{\textstyle{{(|\alpha |c_\theta  /\sqrt 2 )^n } \over {n!}}}} \right)^2 } \sum\limits_{q = 0}^n {{\textstyle{{n!( - 1)^q } \over {(n - q)!q!}}}\sum\limits_{s = 0}^n {{\textstyle{{n!( - 1)^s } \over {(n - s)!s!}}}} } }  \\
   {} &\!\! { \times \sum\limits_{w = 0}^{n - q} {{\textstyle{{(n - q)!( - 1)^w } \over {(n - q - w)!w!}}}\left( {{\textstyle{{s_\theta  } \over {c_\theta  }}}} \right)^{q + w} \sum\limits_{x = 0}^q {{\textstyle{{q!} \over {(q - x)!x!}}}\left( {{\textstyle{{c_\theta  } \over {s_\theta  }}}} \right)^x } } }  \\
   {} &\!\! { \times \sum\limits_{y = 0}^{n - s} {{\textstyle{{(n - s)!( - 1)^y } \over {(n - s - y)!y!}}}\left( {{\textstyle{{s_\theta  } \over {c_\theta  }}}} \right)^{s + y} \sum\limits_{z = 0}^s {{\textstyle{{s!} \over {(s - z)!z!}}}\left( {{\textstyle{{c_\theta  } \over {s_\theta  }}}} \right)^z } } }  \\
   {} &\!\! { \times \sqrt {(n - (w + x))!(w + x)!(n - (y + z))!(y + z)!} }  \\
   {} &\!\! { \times |n - (w + x)\rangle \langle n - (y + z)| \otimes |w + x\rangle \langle y + z|.}  \\
\end{array}
\label{eq:C.1}
\end{equation}
%===============================================================================

Next, we wish to find the single-beam vacuum probability of \smash{$\widetilde{\widetilde{\rho}}$} from (\ref{eq:C.1}), which is \smash{${\widetilde{\widetilde{p}}_{0}}\!\!\!\!{~}^{(1)}\equiv \langle 0|\text{tr}_{2}(\widetilde{\widetilde{\rho}})|0\rangle$}, which introduces Kronecker deltas $\delta_{w+x,k}$, $\delta_{y+z,k}$, $\delta_{w+x,n}$, $\delta_{y+z,n}$, where $k$ is from the sum of the partial trace.  These deltas effectively also cause $\delta_{k,n}$, which picks $k=n$ out of the $k$ sum, and lets us begin by picking $x=k-w=n-w$ and $z=k-y=n-y$ out of the $x$ and $z$ sums.

However, an important point to notice, for example, is that since the limits of the $z$ sum are $z=0$ and $z=s$, then when we pick $z=n-y$ out of that sum, this causes the limit values to yield equations $n-y=0$ and $n-y=s$, which if solved for $y$ produces a restriction on allowable values of $y$ to be $n-s\leq y\leq n$.  But the existing sum on $y$ already places the restriction that $0\leq y\leq n-s$, therefore the only value of $y$ that is allowed between these two restrictions is $y=n-s$, so we see that we have an effective $\delta_{y,n-s}$.  Similarly, we also get an effective $\delta_{w,n-q}$.

Thus, applying deltas within \smash{${\widetilde{\widetilde{p}}_0}\!\!\!\!{~}^{(1)}$} in the order, $\delta_{k,n}$, $\delta_{x,n-w}$, $\delta_{z,n-y}$, $\delta_{w,n-q}$, and $\delta_{y,n-s}$, we arrive at
%===============================================================================
\begin{equation}%                  Equation C.2
\begin{array}{*{20}l}
   {\widetilde{\widetilde{p}}_0^{(1)}  = } &\!\! {e^{ - |\alpha |^2 } \sum\limits_{n = 0}^\infty  {\left( {{\textstyle{{\left( {|\alpha |^2 c_\theta  ^2 /2} \right)^n } \over {\left( {n!} \right)^2 }}}} \right)\left( {{\textstyle{{s_\theta  ^2 } \over {c_\theta  ^2 }}}} \right)^n n!} }  \\
   {} &\!\! { \times \left( {\sum\limits_{q = 0}^n {{\textstyle{{n!} \over {(n - q)!q!}}}\left( {{\textstyle{{c_\theta  } \over {s_\theta  }}}} \right)^q } } \right)^2 \!\!,}  \\
\end{array}
\label{eq:C.2}
\end{equation}
%===============================================================================
where we used the fact that the $q$ and $s$ sums are identical to combine them into a squared sum.  Finally, inserting a $1^{n-q}$ in the squared sum, we recognize it as a squared binomial to power $n$ which then simplifies nicely as
%===============================================================================
\begin{equation}%                  Equation C.3
\begin{array}{*{20}l}
   {\widetilde{\widetilde{p}}_0^{(1)} } &\!\! { = e^{ - |\alpha |^2 } \sum\limits_{n = 0}^\infty  {{\textstyle{{\left( {|\alpha |^2 s_\theta^2 /2} \right)^n } \over {n!}}}\left[ {\left( {1 + {\textstyle{{c_\theta  } \over {s_\theta  }}}} \right)^n } \right]^2 } }  \\
   {} &\!\! { = e^{ - |\alpha |^2 } \sum\limits_{n = 0}^\infty  {{\textstyle{{\left( {{\textstyle{{|\alpha |^2 } \over 2}}(1 + s_{2\theta } )} \right)^n } \over {n!}}}} }  \\
   {} &\!\! { = e^{ - {\textstyle{{|\alpha |^2 } \over 2}}(1 - s_{2\theta } )} ,}  \\
\end{array}
\label{eq:C.3}
\end{equation}
%===============================================================================
which is the result we seek; the vacuum probability of one of the output beams from a VBS that has acted on a joint-beam input \smash{$\rho\equiv B_{45^{\circ}}(\rho_{\alpha_{\overline \varphi}}\otimes\rho_0) B_{45^{\circ}}^{\dag}$} which is the output of a 50:50 BS acting on the product state $\rho_{\alpha_{\overline \varphi}}\otimes\rho_0$, where $\rho_{\alpha_{\overline \varphi}}$ is a Poisson state and $\rho_0 \equiv |0\rangle\langle 0|$.

Since (\ref{eq:C.3}) is the vacuum probability that would be measured at the detector after the VBS after the original laser field has completely phase-damped to a Poisson state, but is equal to that same quantity for the coherent-input case, this shows that the method in \Fig{5} is useless as a test for laser field purity.
%                                  End App.C
%-------------------------------------------------------------------------------
\end{appendix}
%                                END of APPENDIX
%*******************************************************************************
%*******************************************************************************
%                                 BIBLIOGRAPHY
%
%                             END of BIBLIOGRAPHY
%*******************************************************************************
\end{document}